\documentclass[preprint]{iucr}              
\usepackage{amssymb}
\usepackage[fleqn]{amsmath}
\usepackage{graphicx}
\usepackage{tabularx}
\usepackage{booktabs}
\usepackage{array}
\DeclareMathAlphabet{\mathcalligra}{T1}{calligra}{m}{n}
\def\mathbi#1{\textbf{\em #1}}
\numberwithin{equation}{section}
\def\mathbi#1{\textbf{\em #1}}
\DeclareRobustCommand{\captionpar}{\par}
\usepackage{color}
\usepackage{ulem}
\usepackage{url}
\usepackage{yfonts}
\usepackage{relsize} 
\usepackage{wrapfig}

\newcommand{\SVI}[0]{$\bf{S^{6}}$}
\newcommand{\GVI}[0]{$\bf{G^{6}}$}
\newcommand{\CIII}[0]{$\bf{C^{3}}$}

\newcommand{\vdotv}[2]{${{\bf #1 \cdot #2}}$}

\journalcode{A}              
\makeatletter
\font\dummyft@=dummy \relax
\makeatother

\begin{document}                  
	\raggedbottom
	\setlength{\parskip}{0pt}
	{\LARGE \emph{\today}} \\
	\title{Measuring Lattices}
	\shorttitle{Measuring Lattices}
	

	\cauthor[a]{Lawrence C.}{Andrews}{lawrence.andrews@ronininstitute.org}{}
	\author[b]{Herbert J.}{Bernstein}
	
	\aff[a]{Ronin Institute, 9515 NE 137th St, Kirkland, WA, 98034-1820 \country{USA}}
	\aff[b]{Ronin Institute, c/o NSLS-II, Brookhaven National Laboratory, Upton, NY, 11973-5000 \country{USA}}
	
	
	\shortauthor{Andrews and Bernstein}
	
	
	
	
	
	
	
	\maketitle                        
	\begin{synopsis}
        Unit cells are used to represent crystallographic lattices.
		Calculations measuring the differences between unit
		cells are used to provide metrics for measuring meaningful
		distances between three-dimensional crystallographic
		lattices.  This is a surprisingly complex and
		computationally demanding problem.  We present a
		review of the current best practice using 
		Delaunay-reduced unit cells
		in the six-dimensional real space of Selling
		scalar cells \SVI{} and the equivalent three-dimensional complex space \CIII{}.
	\end{synopsis}
	\newcommand{\si}[0]{$s_1$}
	\newcommand{\sii}[0]{$s_2$}
	\newcommand{\siii}[0]{$s_3$}
	\newcommand{\siv}[0]{$s_4$}
	\newcommand{\sv}[0]{$s_5$}
	\newcommand{\svi}[0]{$s_6$}
	\newcommand{\Svec} [0] {\{\si, \sii, \siii, \siv, \sv, \svi \}}
	\newcommand{\SvecA} [0] {\{-\si, -\si+\sii, \si+\siii, \si+\sv, \si+\siv, \si+\svi \}}
	
	\newcommand{\OPES}[0]{$E^3toS^6$}
	\newcommand{\OPESS}[0]{$$E^3toS^6$$}
	\newcommand{\MSVI}[0]{$M_{S^{6}}$}
	\newcommand{\MEIII}[0]{$M_{E^{3}}$}
	\newcommand{\Plus}[0]{$\textfrak{P}$}	
	\newcommand{\Minus}[0]{$\textfrak{M}$}
	
	\newcommand{\ci}[0]{$c_1$}
	\newcommand{\cii}[0]{$c_2$}
	\newcommand{\ciii}[0]{$c_3$}

	\begin{abstract}
		
		Unit cells are used to represent crystallographic lattices.
		Calculations measuring the differences between unit
		cells are used to provide metrics for measuring meaningful
		distances between three-dimensional crystallographic
		lattices.  This is a surprisingly complex and
		computationally demanding problem.  We present a
		review of the current best practice using 
		Delaunay-reduced unit cells
		in the six-dimensional real space of Selling
		scalar cells \SVI{} and the equivalent three-dimensional complex space \CIII{}.
		The process is a simplified version of the process
		needed when working with the more complex
		six-dimensional real space of Niggli-reduced unit
		cells \GVI{}.  Obtaining a distance begins with identification of the fundamental region in the space, continues with
		conversion to primitive cells and reduction, analysis of
		distances to the boundaries of the fundamental unit,
		and is completed by a comparison of direct paths to
		boundary-interrupted paths, looking for a path of
		minimal length.

	\end{abstract}
	
	
	\section{History}
	
	Human fascination with crystals has a long history. 105,000 years ago,
	someone had a collection of calcite crystals 
	(Iceland spar) \cite{wilkins2021innovative}. 
	Theophrastus (ca. 372-287 B.C.), a student of Plato and successor to Aristotle, wrote the first known treatise on gems ("On Stones")  \cite{enwiki:1114534722}.
	
\begin{figure}
	\includegraphics[width=\textwidth]{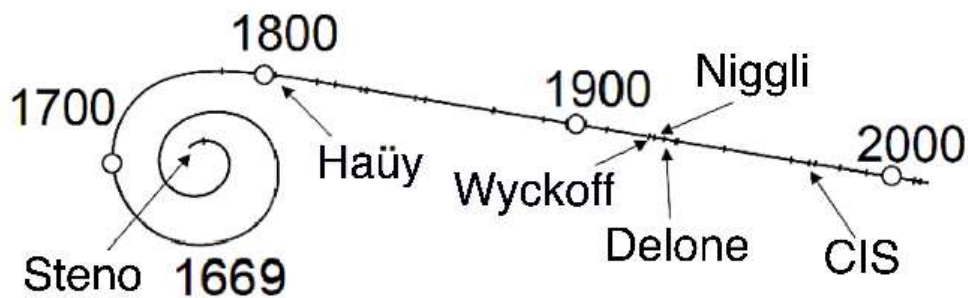}
	\label{timeline}
	\caption{Some key dates in the history of modern crystallography}
\end{figure}

	Figure \ref{timeline} notes a few key events in cataloging 
	crystal properties. We start with \citeasnoun{Kepler1611} (translated in \citeasnoun{kepler1966}) and Steno  (see \citeasnoun{authier2013early}) who
	conjectured on the structures of crystals.  \citeasnoun{hauy1800addition} created the first catalog of minerals.
	
	With the creation of the 1978 version of the Chemical Information System (CIS)  \cite{bernstein1979nih},
	building on the earlier CIS \cite{feldmann1972application} which lacked cell search capabilities,
	online computerized retrieval systems become
	available for searching on ranges of unit cell parameters from
	the Cambridge Structural Database (CSD) \cite{kennard1977computer}. From 
	a few tens of thousands in 1979, the community has progressed to 
	collection of several hundreds of thousands of sets of
	unit cell parameters, with approximately one million in the
	CSD by 2019 \cite{taylor2019million} alone.  When placed in
	a common SAUC database (see below), the number of cells in the 
	Protein Data Bank and the Crystallographic Open Database total
	more than 600,000 entries as of summer 2022.   Below, we discuss algorithms for 
	robust searching on ranges of unit cell parameters. The need for
	ranges can be due to uncertainty of measurements, misidentified
	symmetry classes, variable composition, and searches for
	related materials.
	
	There have been multiple databases that allow searching
	on the values of unit cell parameters, including:
		\begin{itemize}
			\label{databases}
		
		\item Chemical Information System (CIS) (1978 version),
		 transferred to private management in 1984 \cite{kadec1985transfer}.
		 
		\item Protein Data Bank, Rutgers Univ. \cite{Bernstein1977} \cite{Berman2000}.
		
		\item Cambridge Structural Database.  The WebCSD 
		\cite{hayward2019cambridge}	cell search uses the
		\cite{hayward2019cambridge} version of \GVI{}.
		
		\item American Mineralogist Crystal Structure Database \cite{downs2003american}.
		
		\item Crystal Lattice Structures. U.S. Naval Research Laboratory, now in the
		AFLOW library of crytallographic prototypes \cite{mehl2017aflow}.
		
		\item Crystallographic Database for Minerals and Their Structural Analogues, 
		Russian Foundation of Basic Research
		\cite{chichagov2001mincryst}.
		
		\item ICSD Web: the Inorganic Crystal Structure Database, FIZ Karlsruhe
		\cite{ruhl2019inorganic}.
		
		\item Crystallography Open Database, University of Cambridge \cite{gravzulis2012crystallography} 
		
		\item  SAUC (Search of Alternate Unit Cells),  \cite{McGill2014}
	\end{itemize}

Various paradigms are used 
by different databases, sometimes only simple ranges on unit cell
edge length. SAUC uses the methods described herein.	
	
\section{Introduction}

Crystallographers, in general, seem convinced that unit cells
are simple; after all, we learn about them on the first day of
studying crystallography.

Likewise, lattices are considered to be simple. Of course
there are complications, such as space groups, but the concept
of the simple lattice is felt to be almost as simple as
unit cells.
	
Stumbling blocks arise when the need to compare various unit
cells is included in the tasks. For example, if the variations are due
to temperature changes or site substitutions, it is not
uncommon for the variations to be largely confined to a single
parameter, such as one unit cell axial length. 

A different issue arises with the need to compare the unit
cells of an \textbf{a}rbitrary group of materials. For instance, one
might want to search in a database of all known materials. In
those cases, more sophisticated tools are needed. 

What modern problems require comparing arbitrary groups of cells?
\begin{itemize}
	\item database searches (approximately a million unit cells are known)
	\item clustering for serial crystallography (thousands to hundreds of thousands of diffraction images in a single study)
	\item finding candidate materials for solving protein crystal structures 
	\cite{nanao2022id23}
	\item searches for minerals or alloys (both of which may have large compositional variation)
	\item Bravais lattice determination
	\item Studies in epitaxy \cite{yang2014unit}
\end{itemize}

\section{Techniques used to measure lattices}

Several techniques other than common linear algebra methods will be used in measuring distances between unit cells.
\begin{itemize}
	\item 	Metric vector spaces, see Section \ref{spaces}
	\item 	Simple topology, see Section \ref{topology}
	\item   ``Fundamental region" (also called ``fundamental unit"),
	see Section \ref{fundamental}
	\item 	Unit cell reduction, see Section \ref{reduction}
	\item 	Projectors, see Appendix \ref{projectors}

\end{itemize}	

\section{Unit cell parameters}

\label{unitcells}
If we are to measure the distance between two lattices, we will need
to have a space where the measure can be performed. The conventional 
representation of a unit cell is the ``unit cell parameters", 
a, b, c, $\alpha$, $\beta$, $\gamma$. Mathematically, this is a
point in $\bf{R^3 \times R^3}$. That is the product of two different spaces; a
point in one is not necessarily a point in the other; worse in this
case is the fact that a point in the length space means nothing
in the angle space. It is still possible to define a distance, but 
that would require inventing a unit that mixes the units of the length
space and the angle space.
		
\section{Spaces}
\label{spaces}

In order to measure a meaningful distance between two objects, the objects need to
be placed in a ”metric space”. For crystallographic unit cells, this will often be a
vector space amenable to the tools and techniques of
linear algebra, at least locally. Because six free parameters are needed to define a unit
cell, the space will need to be at least six-dimensional. Depending on
other criteria, more dimensions may be needed; gluing the edges of
a manifold generates the need for more dimensions.

\section{A digression on topology}
\label{topology}

One aspect of topology is the study of objects and the relationships
of their edges, which can be ``glued" together to make new objects
and disclose new relationships among the parts. Here we look at a simple 
example that demonstrates some of the issues that arise
in measuring distances between lattices.

\subsection{Create a simple, 1D unit cell}
	
\begin{figure}
	\includegraphics[width=\textwidth]{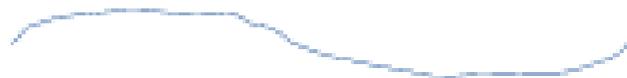}
	\label{line_1}
	\caption{This is just an empty single unit cell.}
\end{figure}

\subsection{Put two atoms into the unit cell}
	
	Here we put atoms at the ends of the cell so that they
	will be close to their counterparts in the adjacent cell.
\begin{figure}
	\includegraphics[width=0.4\textwidth]{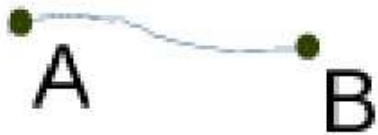}
	\label{line_2}
	\caption{Put an atom at or near each of two boundaries of the unit cell}
\end{figure}

\subsection{Connect the unit cell with its neighbors}
	
	Here we see that the ``correct" distance between A and B is
	quite small compared to the distance within the 
	fundamental cell.
\begin{figure}
	\includegraphics[width=\textwidth]{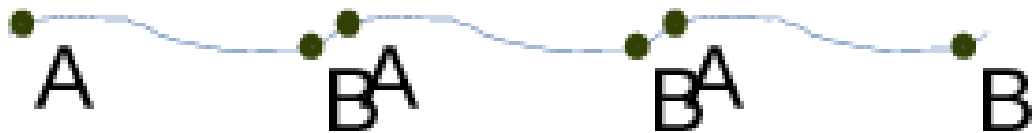}
	\label{line_3}
	\caption{Here we show just the nearest neighbor cells}
\end{figure}

\subsection{Another choice: wrap around into a loop}
	A technique based in topology is to wrap one edge of a
	cell through a higher dimensional space to join
	another edge of the cell to make a loop. Doing this can make the 
	problem seem simpler, since there is only one cell
	to be considered. However, the problem of distance calculation has
	not been simplified. Now the distance has to be measured
	in both directions around the loop, and the shorter distance
	reported. This process of joining is termed ``gluing". In
	addition, the distance must be measured within the curved
	surface.
	
     Gluing two edges of the one-dimensional cell
	turns the cell into a circle, making the problem
	two dimensional. 	If 
	we had started with a two-dimensional cell,
	then gluing opposite pairs of edges could be used to
	make a three-dimensional torus. In either case, the dimensionality of the system
	would be increased by one. 
\begin{figure}
	\includegraphics[width=\textwidth]{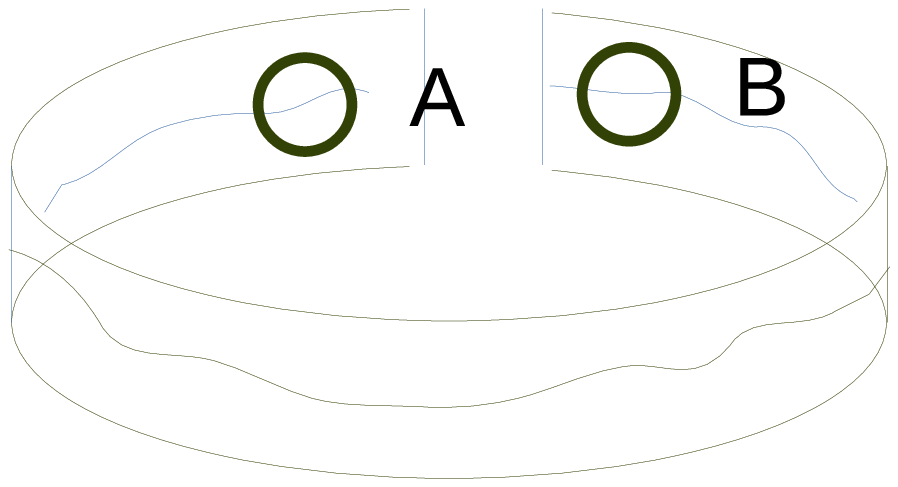}		
	\label{loop}
	\caption{}
\end{figure}

\subsection{Or you could twist the loop to make a M{\"o}bius strip}

Consider the above case where the cell was on a 2-D surface.
Instead of simply joining the ends, we can twist the loop before
we glue the ends together, making a M{\"o}bius strip.


\begin{figure}
	\includegraphics[height=0.45\textwidth]{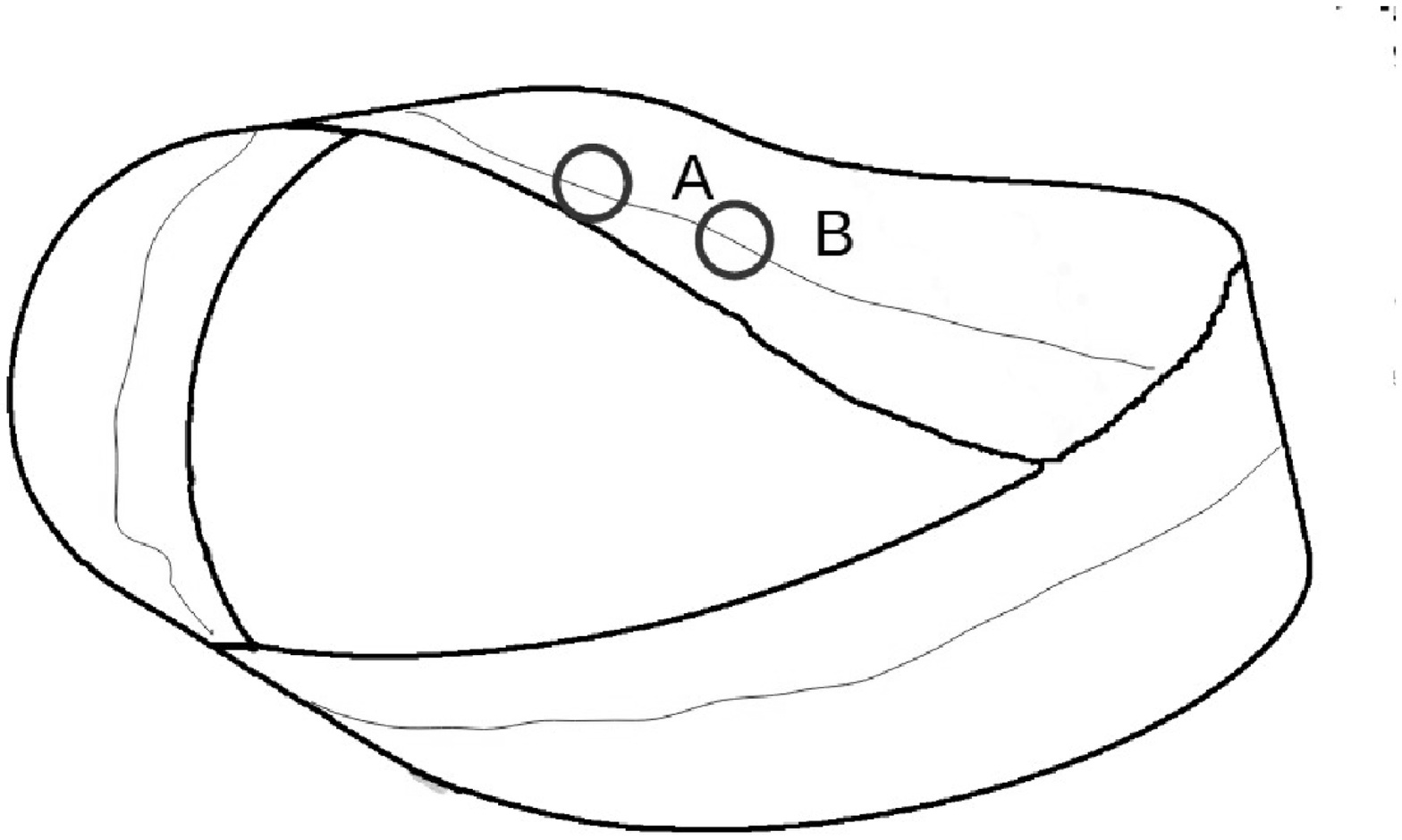}
	\label{mobius_1}
	\caption{Twisting before joining makes a M{\"o}bius strip}
\end{figure}

\subsection{Summary of the properties on joining}

One aspect of gluing is that the dimensionality increases. The issue of measuring distances also 
become more complex. With added dimensions, the directions in which to look for shortest paths may increase. Similarly
the glued boundaries themselves add to the cases to
be considered in looking for shortest paths.
		
\section{Spaces, continued}

In this paper, we use several spaces to represent unit cells.
Other choices of spaces are possible, but this set will suffice 
for the current topics.
		
\begin{itemize}
	\item 	Lengths in \r{A}ngstroms and angles in degrees, a 
	space of $R^3 \times R^3$).
Although the presentation of unit cell parameters as edge lengths and
interaxial angles is familiar, it does not present an obvious metric (see Section \ref{unitcells}). A
possible alternative would be to treat each length and angle pair as
a point in polar coordinates. This unexplored variation makes use
of the association of each axis with its related angle,\textit{ e.g.} \textbf{a}
and $\alpha$ (see appendix \ref{polar}).
	\item 	\GVI{} (Niggli space)
	(see appendix \ref{G6})
	\item 	\SVI{} (Delone space)
	(see appendix \ref{S6})
	\item 	\CIII{} 
	(see appendix \ref{C3})
\end{itemize}	

For future reference, we are in the process of extending 
this list by consideration of a seven-dimensional space of Wigner-Seitz cells \cite{Bernstein2023}.
	

	\section{Fundamental region}
	\label{fundamentaltext}
	
	In working with a space in which there are 
	multiple symmetry-related occurrences of items of interest (atoms, molecules, or,
	in this case, unit cells), it is conventional to choose a fundamental
	region \cite{enwiki:1028832752}. In a two-dimensional graph, it would often be chosen as the all-plus
	quadrant. In crystallography, it might be a unit cell, a
	particular asymmetric unit of a crystal, or even a molecule.
	
	The conventional fundamental unit for Niggli space is the region
	of \GVI{} that contains the Niggli-reduced unit cells. This choice
	has the disadvantage that the fundamental region is non-convex,
	and the boundaries are complex \cite{Andrews2014}.
	
	For Delone space, the fundamental unit is conventionally chosen
	as the all-negative orthant of \SVI{}, which contains all 
	reduced cells. In this orthant, all non-boundary 
	points (that is, those having no zero scalars) correspond to
	a computable unit cell. All points with one or two zero scalars also
	correspond to valid unit cells. Boundaries that have
	four, five, or six zeros have no valid cells. For three zeros, there are
	valid cells except for those points with a prototype of
	$s_1, s_2, s_3, 0, 0, 0$ \cite{Andrews2019b}.
	
	The all-negative orthant (excluding the boundaries) is 
	an ``open" manifold in topological terms; all points
	there correspond to exactly one lattice. The fact that
	certain boundaries do not correspond to valid unit
	cells complicates certain aspects of its topology.

	\section{Measuring}

	Measuring the distance between two lattices
	comes down to finding a geodesic between them, {\it i.e.}
	a minimal length path between them obeying the rules for points
	in the space. We first give a simple example of points in
	a 2-D manifold, which demonstrates some of the complexity of treating lattices.
	
	\subsection{Step 1, fundamental unit}
	\label{fundamental}
	
	The obvious step to find the shortest distance between two
	lattices is to simply compute the Euclidean distance between
	them in the fundamental unit. That distance might not be the
	shortest that can be found, but no geodesic could 
	be longer. It makes a useful starting point. Call 
	this distance $d_0$.
	
	\subsection{Boundaries}
	
	The next step is to consider the boundaries of the fundamental
	unit. If there is no boundary with a shorter distance to
	 \textit{either} of	the lattices than $d_0$, then $d_0$ is the required minimum.  
	 
	If either lattice does have a boundary closer to it than
	$d_0$, call the shortest distance from the first lattice
	to any boundary $d_1$ and the shortest distance
	from the second lattice to any boundary $d_2$.  If 
	$d_1 +d_2 > d_0$ then $d_0$ is still the required
	minimum and we are done.   It is only when both lattices
	are closer than this to some boundaries that we need to
	engage in a more complex search for geodesics. 
	
	As a simple example of treating boundaries, consider a plane with a single
	symmetry element, two-dimensional point group 4.
	
	\includegraphics[width = 0.5\textwidth ]{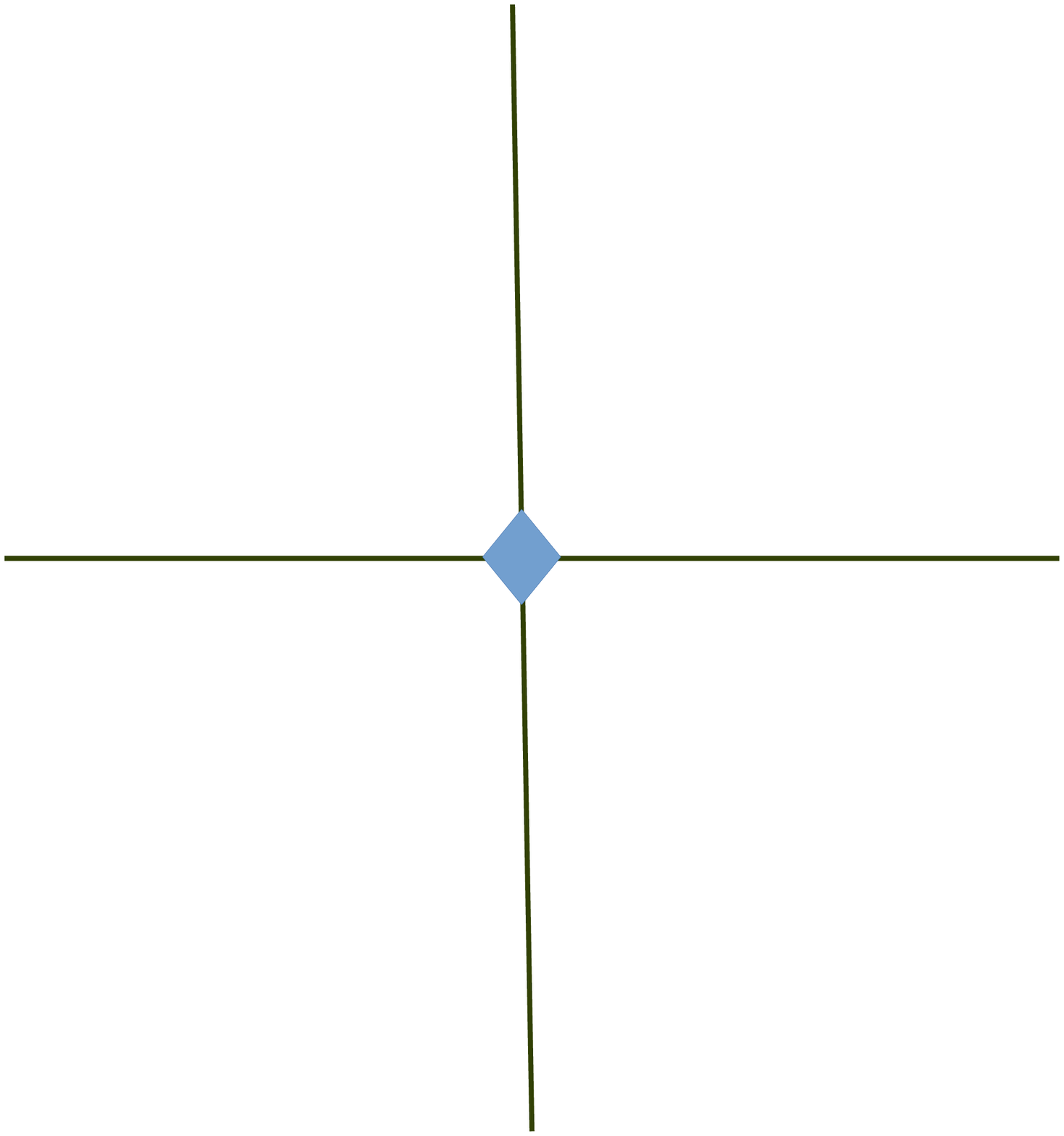}
	
	We choose the lower left quadrant to be the fundamental unit.
    This is an arbitrary choice.
	
	\includegraphics[width = 0.5\textwidth ]{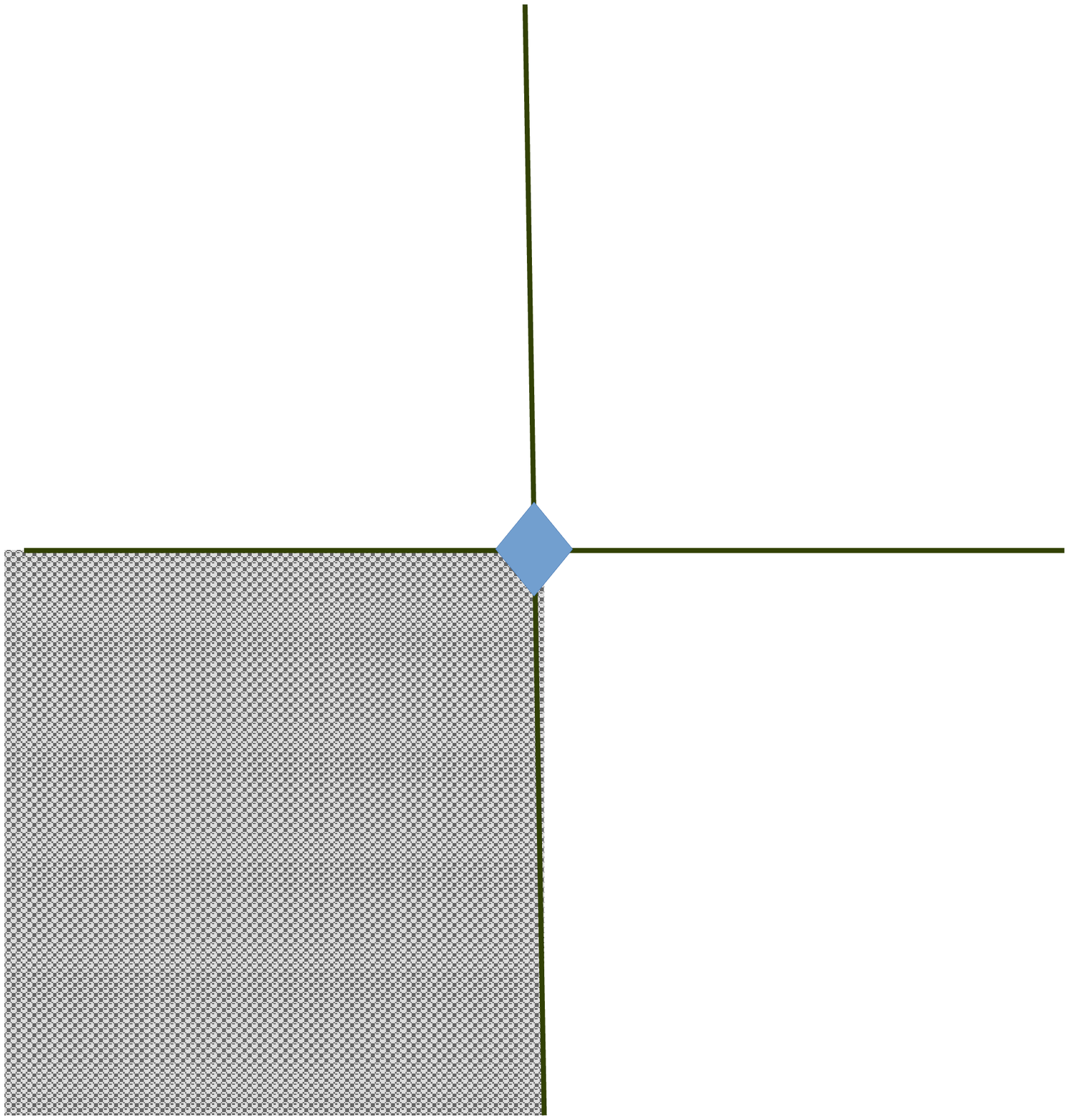}
	
	Now put a single atom, A, in the plane, and apply the
	4-fold operation.
	
		\includegraphics[width = 0.5\textwidth ]{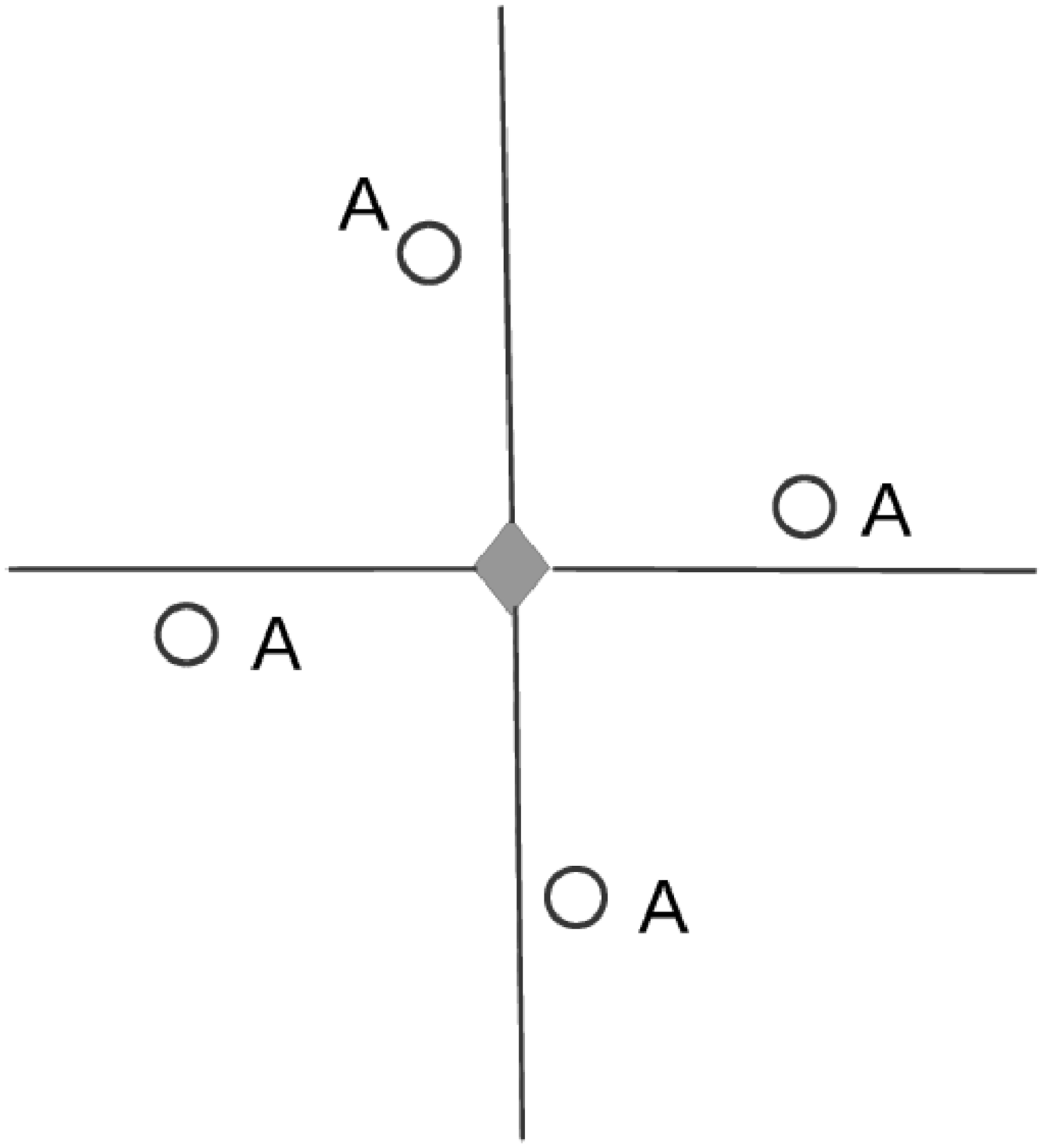}
		
	Now let us consider what happens if we place three atoms, A, B,  and C in the fundamental unit.
	
\begin{figure}
		\includegraphics[width = 0.5\textwidth ]{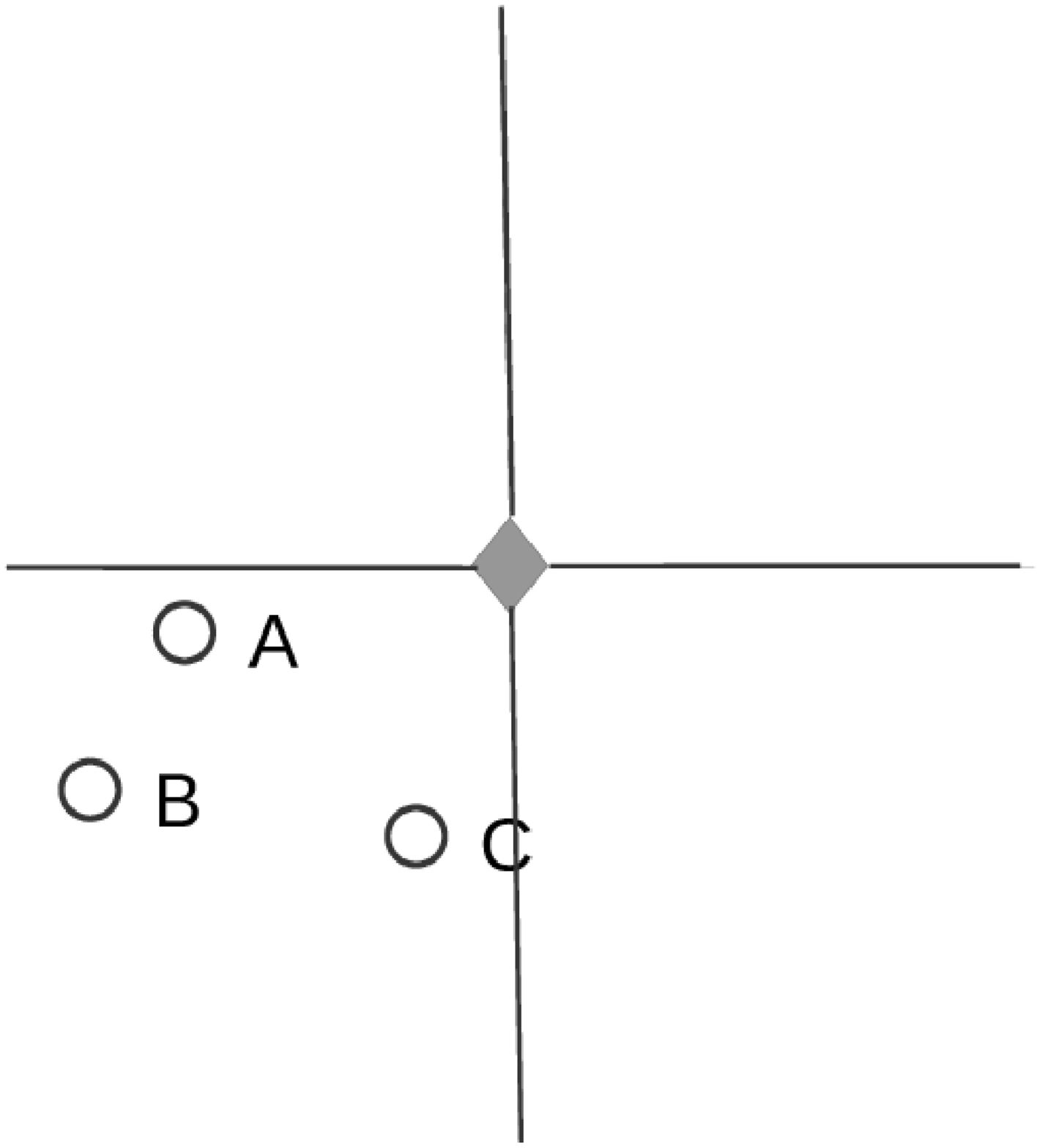}
		\label{4_3}
	\caption{}
\end{figure}	
	It is clear that within the fundamental unit, ignoring all symmetry
	transforms and considering B and C, B is the atom closer to
	A. Draw a circle centered on A that touches B. We see that the
	circle crosses the boundary and a portion is outside of the
	fundamental unit. Any atom that is within that region outside the fundamental unit must be examined to see if it contains
	an atom.
	
	\includegraphics[width = 0.5\textwidth ]{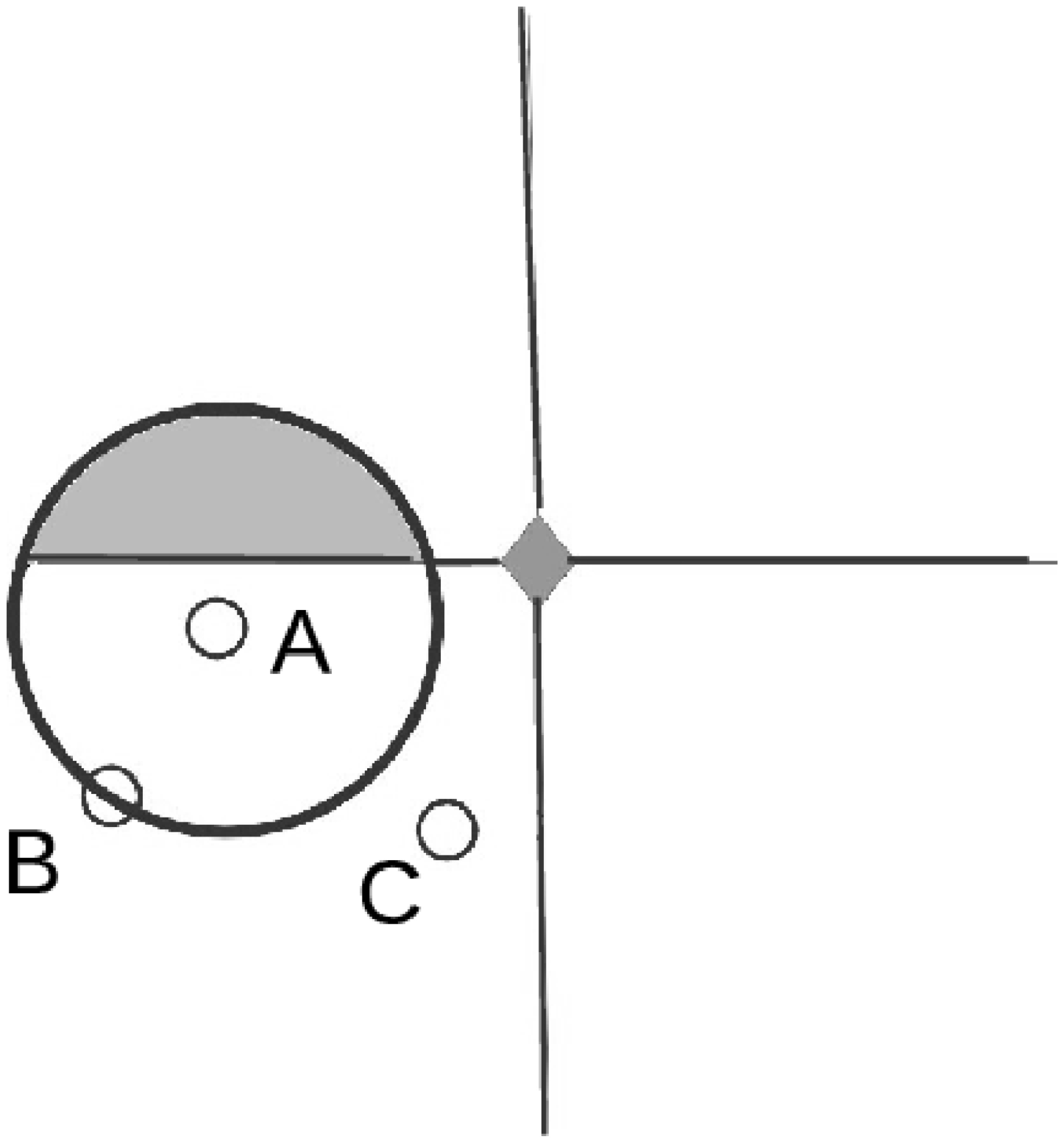}
	
	Applying the 4-fold operation to the shaded segment, we
	see that the atom C (in the fundamental unit) is closer to A
	than B is to A.
	
	\includegraphics[width = 0.5\textwidth ]{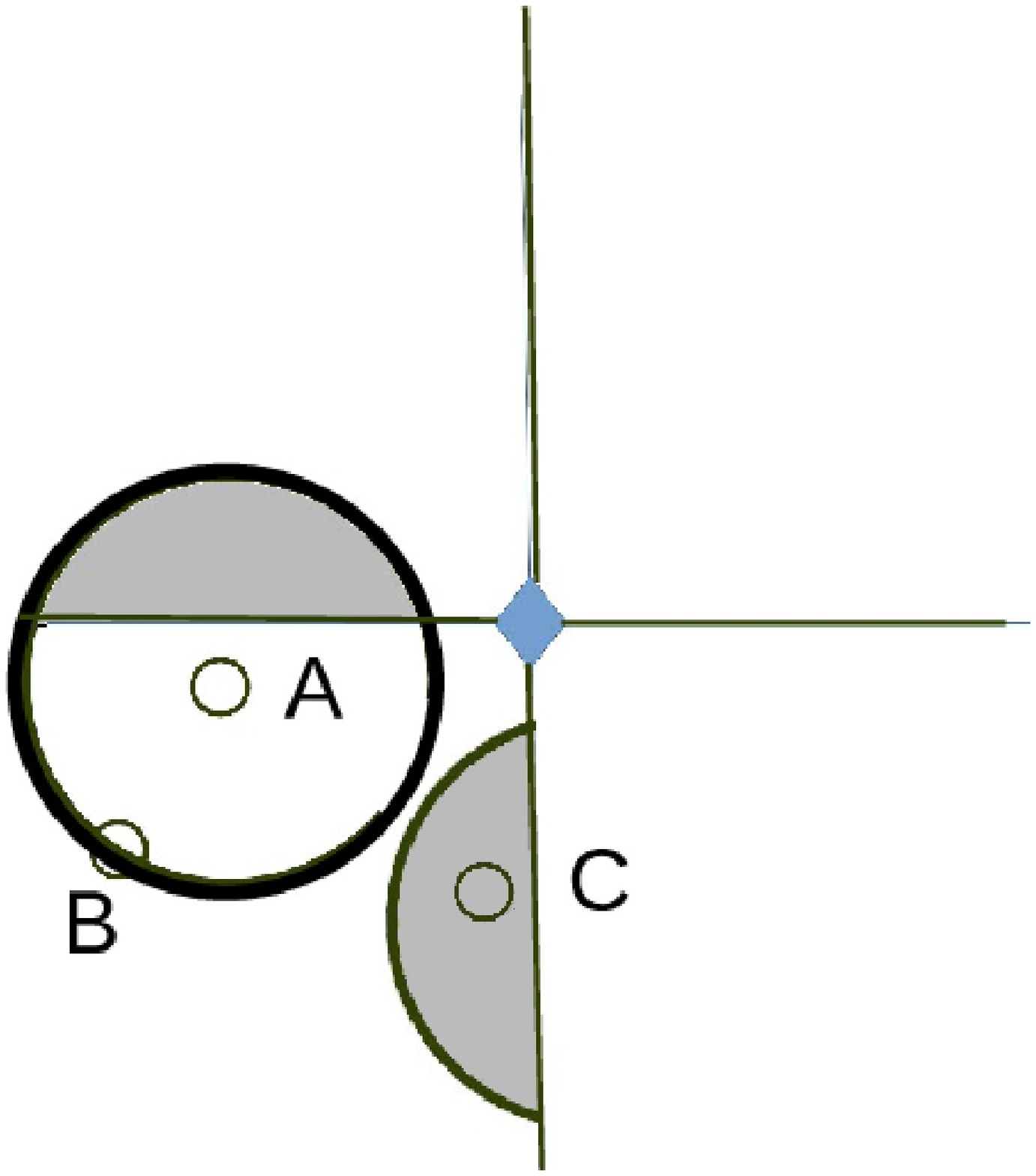}
	
	The next part of the problem is how to compute the actual 
	distance from A to C. The obvious answer is to apply the 
	4-fold symmetry operation either to A or to C and measure
	the Euclidean distance. We will see in the next section that
	such simplistic solutions are not sufficient for lattices.
	
	\subsection{Spaces for measuring}
	
	From this point on, only two spaces will be used for
	specific examples, \SVI{}, the
	space of Selling scalars as six real vector components,
	$( P,Q,R,S,T,U )$ = 
	({$\mathbf{b} \cdot \mathbf{c}$},
	{$\mathbf{a} \cdot \mathbf{c}$},
	{$\mathbf{a} \cdot \mathbf{b}$},
	{$\mathbf{a} \cdot \mathbf{d}$},
	{$\mathbf{b} \cdot \mathbf{d}$},
	{$\mathbf{c} \cdot \mathbf{d}$}) where {$\mathbf{d}=-\mathbf{a}-\mathbf{b}-\mathbf{c}$}, 
	 and \CIII{}, the same scalars as \SVI{}, but used as the real and imaginary parts
	of three complex vector components, $(P+iS, Q+iT, R+iU)$, with each complex number written as a column two-vector.  See sections \ref{S6} and
	\ref{C3} in the appendices. \CIII{} is simply an alter ego of \SVI{}, but 
	some operations are simpler to describe in \CIII{}. The
	reason for limiting our consideration to these two spaces is their simplicity
	compared to other spaces. As stated above, the fundamental
	unit of \SVI{} is a simply-connected, convex manifold, with
	only a few excluded regions (multiple zeros).
	
	For any given lattice there are infinitely many choices of cells from that lattice that can be used to represent it.  The point of reduction is to unambiguously choose a single one of those choices of cells.  In the absence of
	experimental error that is what, for example, Niggli reduction does.  In the presence of experimental error things are not
	so simple and any of several equally valid,
	but very different, unit cells may end up being chosen
	to represent the same lattice \cite{Gruber1973} \cite{McGill2014}.

	\subsection{Reflections, a simple example}
	
	Reflections in \SVI{} were defined in \citeasnoun{Andrews2019b}. They are
	the operations that create isometric copies of a lattice, always within the
	fundamental unit. When comparing two lattices, it is necessary to 
	examine the distances from a reduced cell in one lattice to each of the reflections of another reduced cell in the other lattice.
	
\begin{centering}
\begin{figure}
	\includegraphics[width=.8\textwidth]{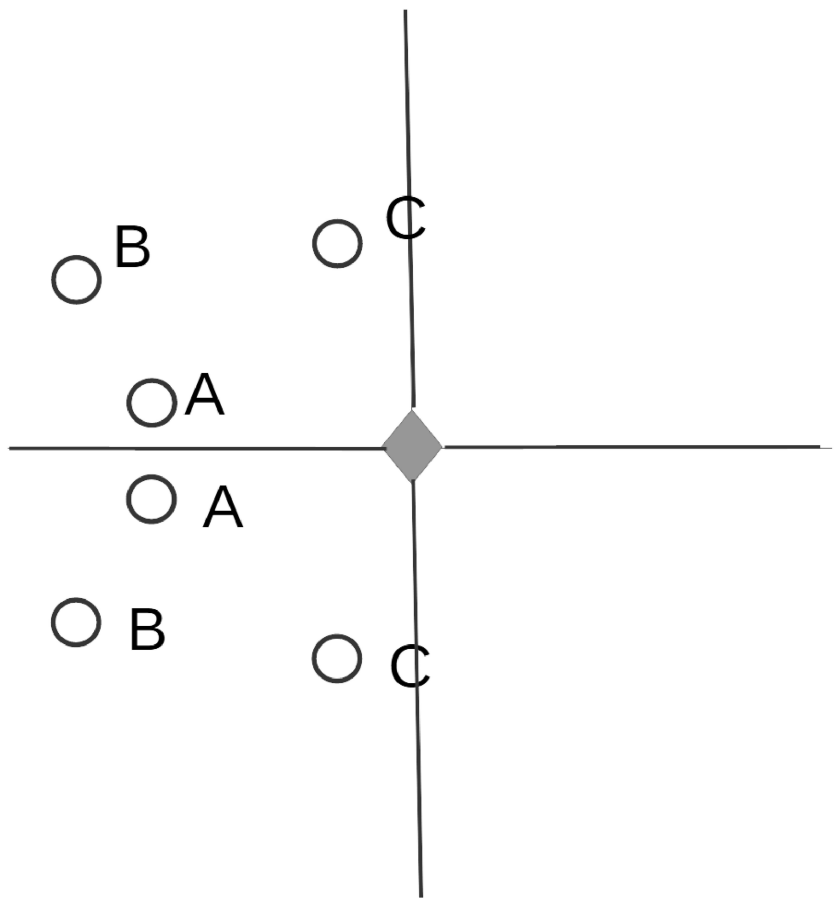}
	\label{4mm_3}
	\caption{Recreating the above example, but with point group 4mm. }
\end{figure}
\end{centering}

	Returning to Figure \ref{4_3} (where the distance to objects
	outside the fundamental unit were considered), consider instead the
	 two-dimensional point group 4mm. Figure
	\ref{4mm_3} shows a subset of the reflections of the basic set of three points.
	Immediately, it is clear that in this case, the 
	distance between two copies of
	point A is the shortest.
	
\subsection{Reflections in \SVI{} and \CIII{}}

``Reflections" were defined by\citeasnoun{Andrews2019b} as
the 24 operations that create different, reduced versions of
a reduced cell in \SVI{} and \CIII{} (isometric copies).
By this definition, the 
4-fold operation and the mirrors in point group 4mm (see Figure \ref{4mm_3})
are reflections.

	An important difference between Niggli space and Delone space
	is that Niggli space has the vector scalars sorted by 
	somewhat complex rules. \SVI{} (that is, Delone space)
	being unsorted results in there being multiple points in the
	fundamental unit for each lattice. In general, there are 24 
	reflections, but equal scalars and zero scalars can lead to degeneracies.
	
	The reflections are simpler to describe in \CIII{}, but,
	of course, the same ones exist in \SVI{}. Here we use the \CIII{} vectors: $c_1$, $c_2$, and $c_3$, see Appendix \ref{C3}.
	
First there are the three interchanges of scalars: $c_1\Longleftrightarrow c_2$, $c_1\Longleftrightarrow c_3$, 
and $c_2\Longleftrightarrow c_3$.

There are also reflections that interchange elements within 
scalars.
\begin{figure}
		\includegraphics[width=0.5\textwidth]{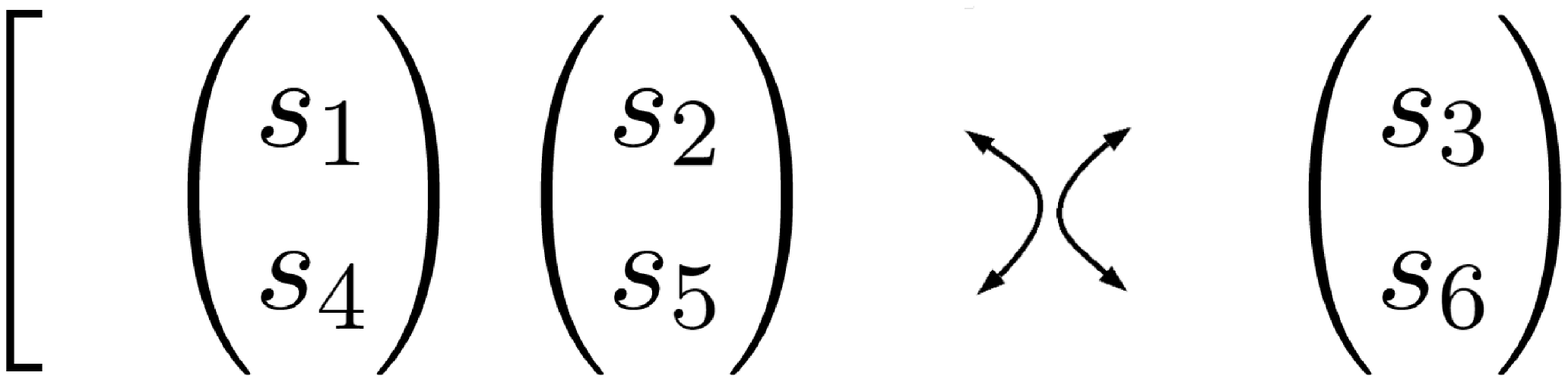}
%
%

	\caption{Example of interchanging real and imaginary scalars
	for the case of reduction of scalar $s_1$.}
	
\end{figure}

And finally each pair of the reflections needs to be combined
giving a total of 24.

\subsection{Reduction}
\label{reduction}

\begin{figure}
	\includegraphics[height=0.65\textwidth]{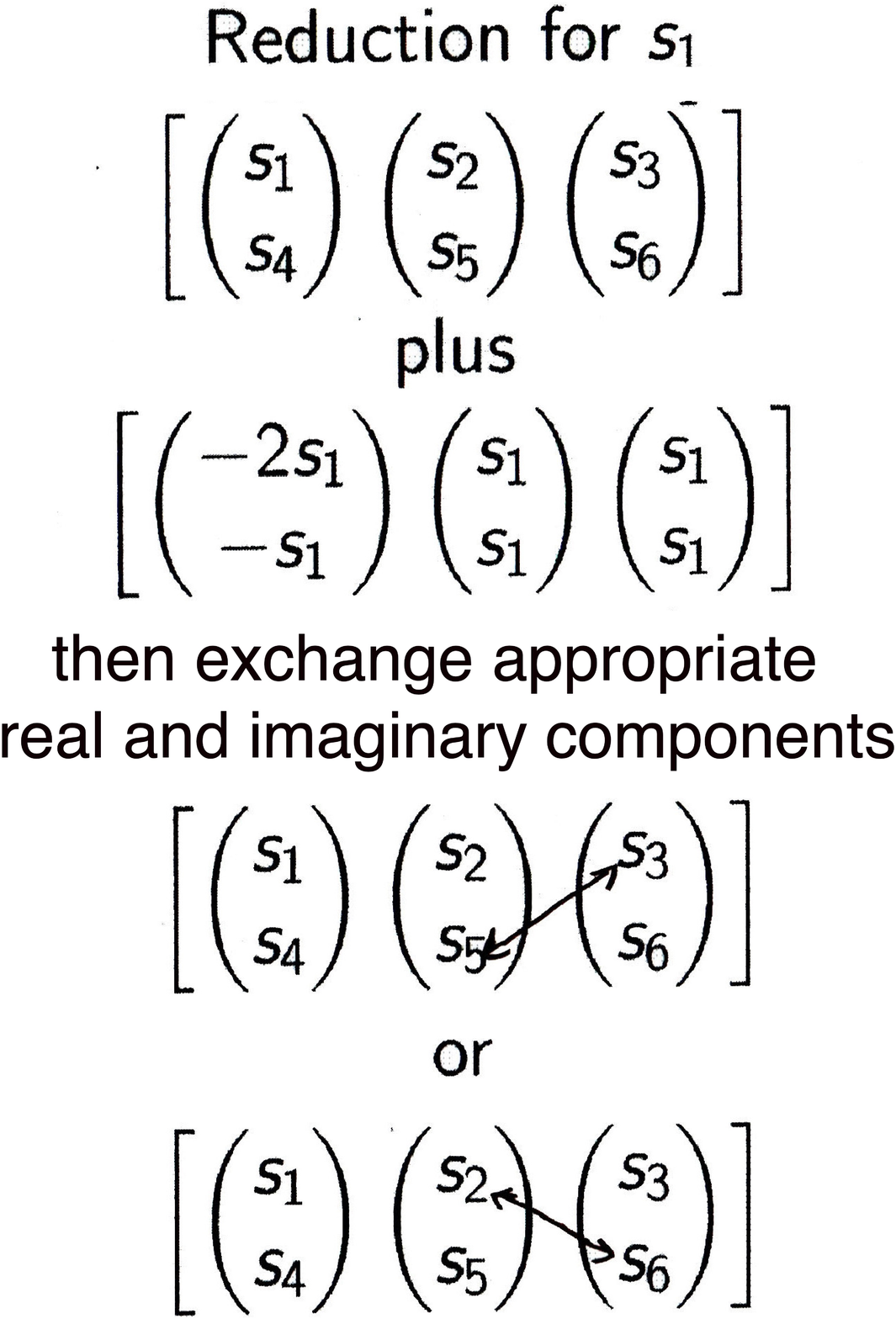}
	\label{S6_ops_B}
	\caption{Steps for reduction in \CIII{}}
\end{figure}

The objective of reduction in \SVI{} (and in \CIII{}) is to
make each Selling scalar less than or equal to zero.
If any scalar is greater than zero, there are two stages
to a reduction step. First,
a vector is composed and added to the vector being reduced.
Then two scalars are interchanged as shown in Figure \ref{S6_ops_B}.

While any scalar can be the one to be reduced, this figure shows
the operations for $s_1$.

	\subsection{Boundary transformations}
	
	The transformations for crossing boundaries of the 
	fundamental unit are the reduction operations.
	
		The section on fundamental regions (see Section \ref{fundamentaltext}) shows the problem of points outside
	the fundamental unit and how boundary transformations were applied
	there. The boundary transformations for lattices are more complex
	than those of space groups. 
	
\begin{enumerate}
	\item  The boundaries must be identified.
	\item  The transformations at the boundaries in \GVI{} and
	 	\SVI{} are not isometric, meaning that they do not preserve
	 	distance measure.
	\item  For a particular boundary, there may be more than one
	 	allowed transformation available.
	\item  Applying a transformation to a point near a boundary may generate
	 	a new point that is near another boundary, requiring further
	 	searching.
	 \item A point that is in the fundamental unit but
	 close to a boundary might need to be transformed in order to
	 find other nearby points.
	 \item All measurements must be done in the metric of the
	 fundamental unit. A corollary is that when two points are
	 being treated, one must be held fixed and within the 
	 fundamental unit while the possible boundary transformations 
	 and reflections of the other are treated.
	 \item If a point is near more than one boundary,
	 then there are multiple transformations and reflections
	  to consider.
\end{enumerate}	

	The fundamental unit in \SVI{} is the all-negative orthant. The
	boundaries are all where one (or more) of the scalars equals zero.
	Two points that are close together and have a corresponding scalar that is positive 
	in one and negative in the other may have representations in 
	the fundamental unit that are far apart. 
	
\begin{figure}
		\includegraphics[width =0.8\textwidth ]{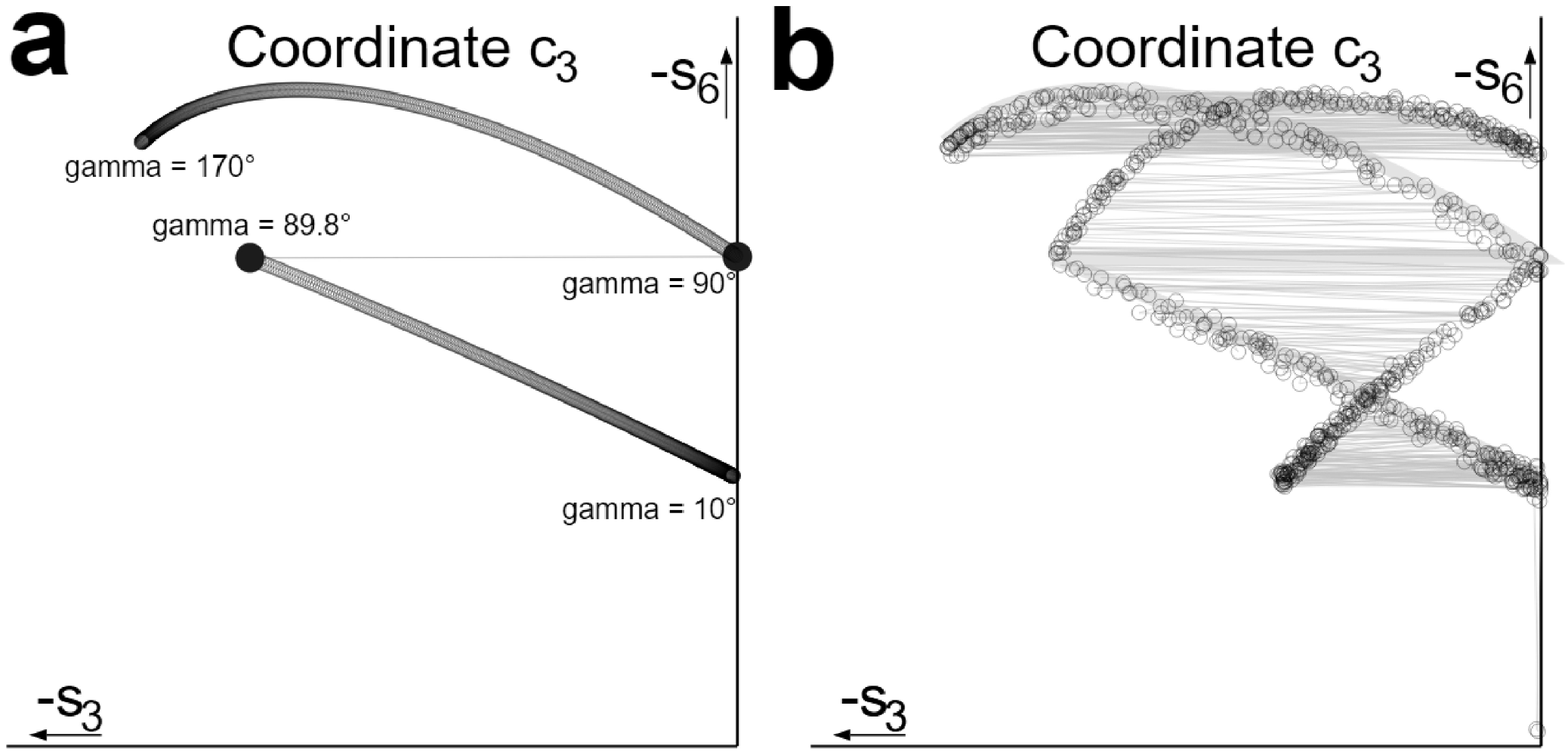}
		\label{DUAL}

	\caption{
	\captionpar{}In both figures (a) and (b), a series of points has been generated in \SVI{}. The basic cell is a=10, b=10, c=10, $\alpha=90^{\circ}$, $\beta=90^{\circ}$, and $\gamma$ ranges sequentially from $10^{\circ}$ to $170^{\circ}$, in $0.2^{\circ}$ increments. Each cell was Delone-reduced. Each was then placed in the \CIII{} asymmetric unit \cite{Andrews2019b}. Then, in sequence from the first cell, for the succeeding cell, all 24 reflections of the succeeding cell were examined, and the reflection that is closest to the first cell chosen. Calculation then proceeds to the next pair (where the new first is the second chosen in the previous step). A line (shown in gray) was drawn between each successive pair of points. The individual points are shown as circles.
	\captionpar{}~~
	\captionpar{}This figure shows the points converted to \protect\CIII, and the c3 coordinate is displayed.  In figure (a) on the left the points are generated as described. The gray connection lines, except for the one connection $\gamma=90^{\circ}$ to $\gamma=89.8^{\circ}$, are hidden by the circles. That single long line represents a case when distance measured in the fundamental unit is quite long, although the originally generated points are quite close together. In figure (b) on the right, initially, three copies of each point were generated, and all points were perturbed by 1\% in a random direction orthogonal to the generated vector (point). Then (in the order of generation), the above sequencing was performed.
    \captionpar{}~~
    \captionpar{}Because the basic structure of the point generation requires that  all points be on multiple boundaries, there are many cases where the distances in the fundamental unit are long. Every gray line is a case where the simple Euclidean distance in the fundamental unit is long, but the generated points are close together.
	}
\end{figure}

	In Figure \ref{DUAL}a,
	the two points with $\gamma=89.8^{\circ}$  and $\gamma=90^{\circ}$ are far apart.
	
	Figure \ref{DUAL}b demonstrates a further complexity. Because the angles $\alpha$
	and $\beta$ were chosen to be $90^{\circ}$, all points in Figure \ref{DUAL}a sit
	on two boundaries ($beta$=$90^{\circ}$ and  $\gamma$=$90^{\circ}$, 
	$s_4$=$0$ and $s_5$=$0$). Perturbing those boundary points may cause them to be within
	the fundamental unit, but, on the other hand,  may cause them to be outside and thus be far from
	the neighboring points (after reduction). All the gray lines
	correspond to members of lattice pairs that are close together
	but are far apart in the fundamental unit (that is, Delone-reduced).
	
\begin{figure}
	\includegraphics[width=0.5\textwidth]{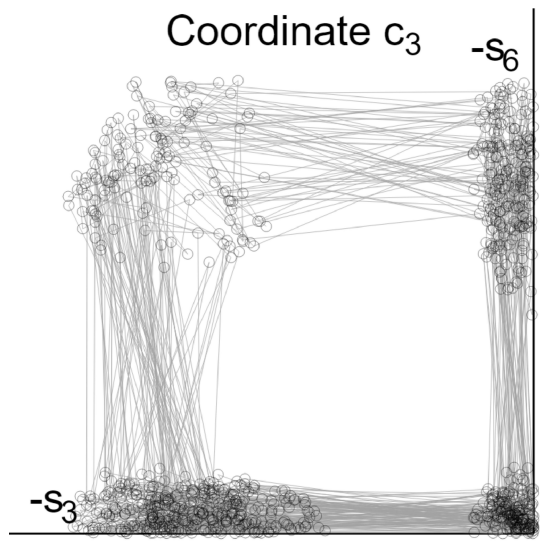}\label{perturbed}

	\caption{1000 copies of \SVI{} vector $\begin{bmatrix}{ 0, 0, 0, 10, 10, 10}\end{bmatrix}$ were 
	perturbed by 1\% orthogonal to the vector. As in Figure \ref{DUAL}, for
	each pair of points, the second point was chosen to be the reflection that
	is closest to the first point. The choice of reflection pairs was done by
	treating the list of generated points as a sequence. Long lines indicate
	pairs that are far apart within the fundamental unit. }
\end{figure}

As seen in Figure \ref{DUAL}, points that are a small distance apart may
be far apart in the fundamental unit. As an example, we choose a primitive
cubic unit cell, \SVI{} vector $\begin{bmatrix}{ 0, 0, 0, 10, 10, 10}\end{bmatrix}$. In Figure \ref{perturbed},
1000 copies have all been perturbed by 1\% in a direction orthogonal to that vector.

\section{Boundaries (and reduction) in \SVI{} and \CIII{}}

	\subsection{Multiple boundaries}
	
	If a point is near two boundaries, all the cases for each boundary need to be considered (recall
	that in \SVI{} there are four possible boundary transformations at
	each of the six boundaries).
	In addition, applying the transformations of one of the boundaries
	and then those of the second is necessary. That means that
	there are 16 possible results. Further still, the 
	case of applying them in the other order must be considered (resulting
	in another 16 cases).
	
	For the case of being near three boundaries, similar complexity to
	the case for two boundaries must be considered: all single 
	boundaries, all boundary pairs (in both orders) and all 
	triples (in all possible orders).
	
	\subsection{Repeated boundaries}
	
	In some cases, applying a boundary transformation results
	in one of the modified scalars being near a boundary
	that it was not near before. Such a case requires that the
	new neighbor boundary also be considered.

	\subsection{Iteration}
	
	Much of the complexity of determining the shortest distance between two
	lattices is due to the need to repeat steps of reduction and reflection.
	Among the places in the process where this becomes necessary is when a
	reduction step creates a new point that is near a boundary.
	
	\subsection{Measuring lattice distances in \SVI{}}

\begin{figure}
\includegraphics[height=0.6\textwidth]{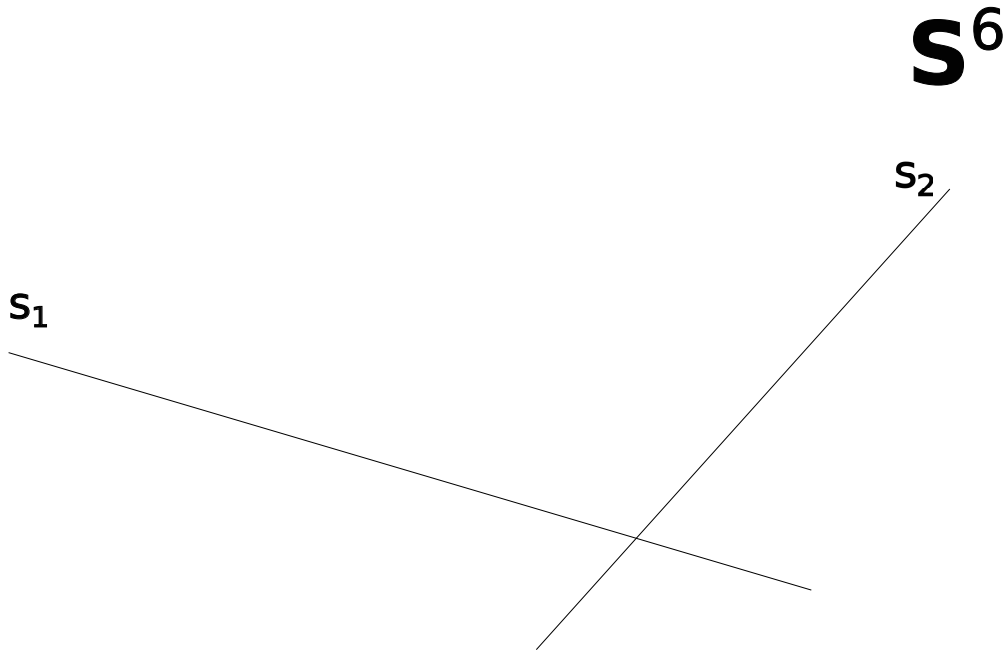}
\caption{In space \SVI{}, choose two axes. The operations shown
will be a cartoon of the operations in six-space.}
\label{S6_1}
\end{figure}

\begin{figure}

\includegraphics[height=0.6\textwidth]{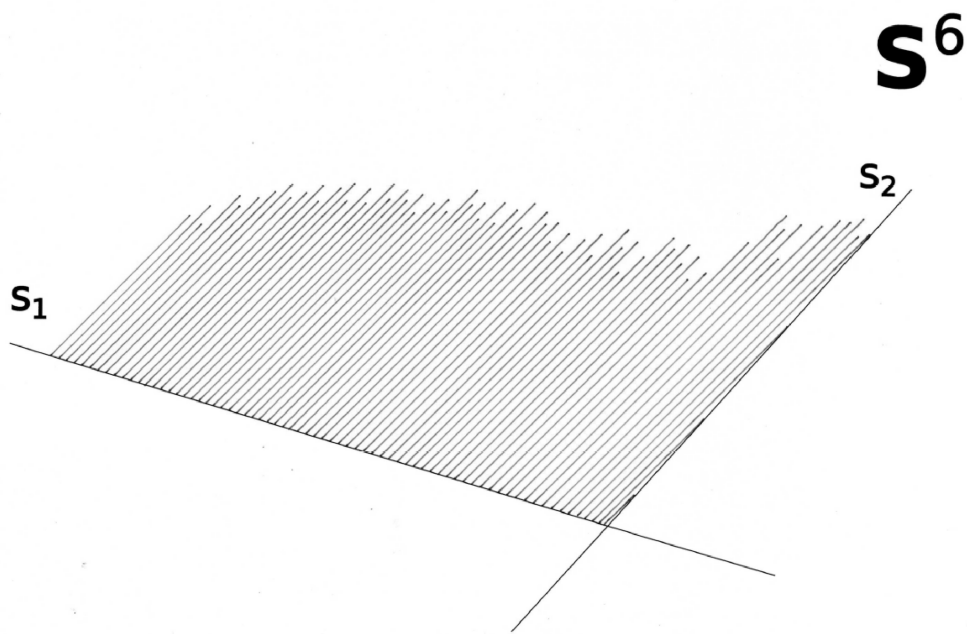}
\caption{Define a fundamental unit (all non-positive region).}
\label{S6_2}
\end{figure}

\begin{figure}
\includegraphics[height=0.6\textwidth]{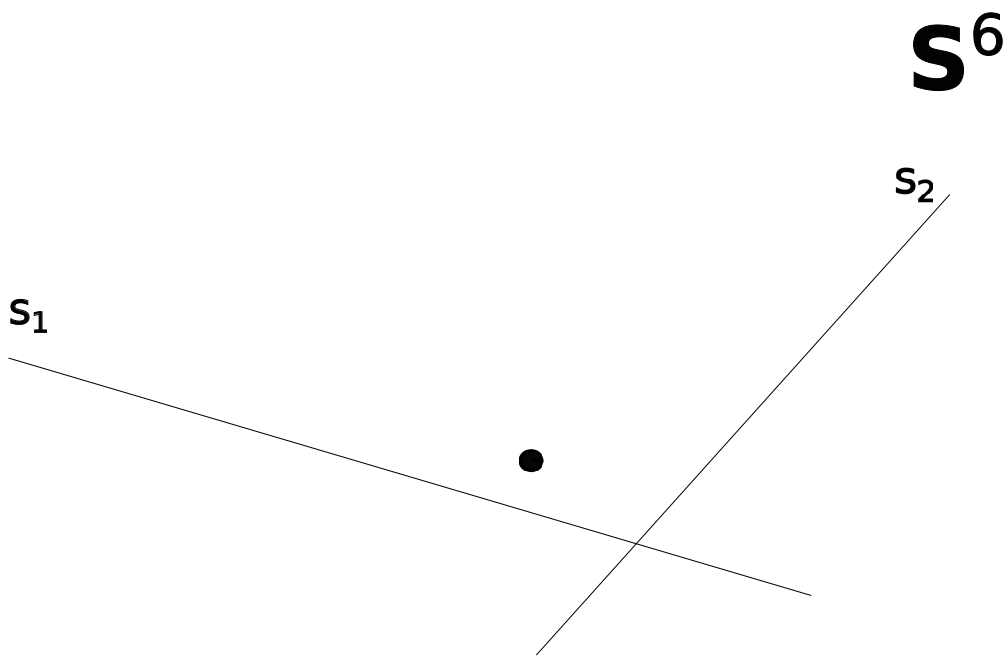}
\caption{Put in a unit cell (a point in \SVI{}).}
\label{S6_3}
\end{figure}

\begin{figure}
\includegraphics[height=0.6\textwidth]{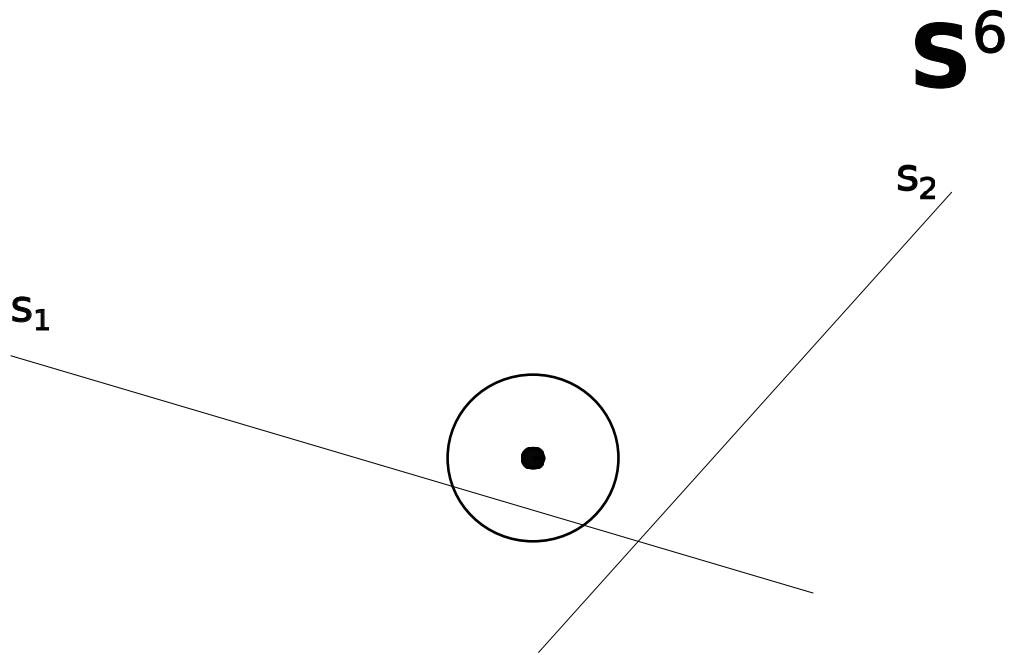}
\caption{Define a search radius. In this case, it
	crosses a boundary and a region is outside the
	fundamental unit. Cells there are \textit{not} reduced.}
\label{S6_4}
\end{figure}

\begin{figure}
\includegraphics[height=0.6\textwidth]{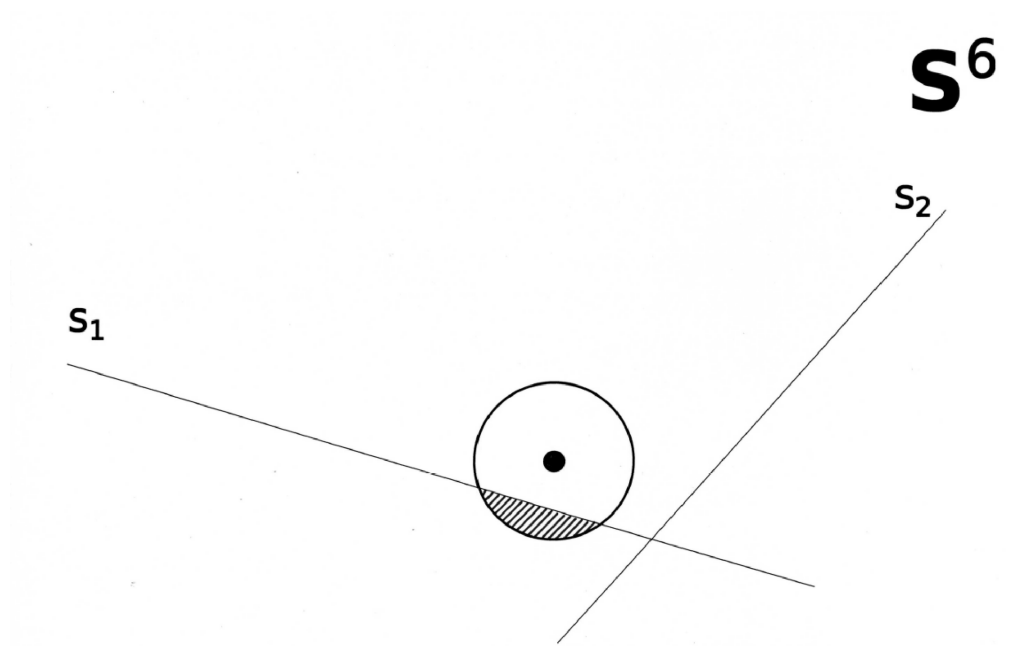}
\caption{The non-reduced region is marked in crosshatch.}
\label{S6_5}
\end{figure}

\begin{figure}
	\includegraphics[height=0.6\textwidth]{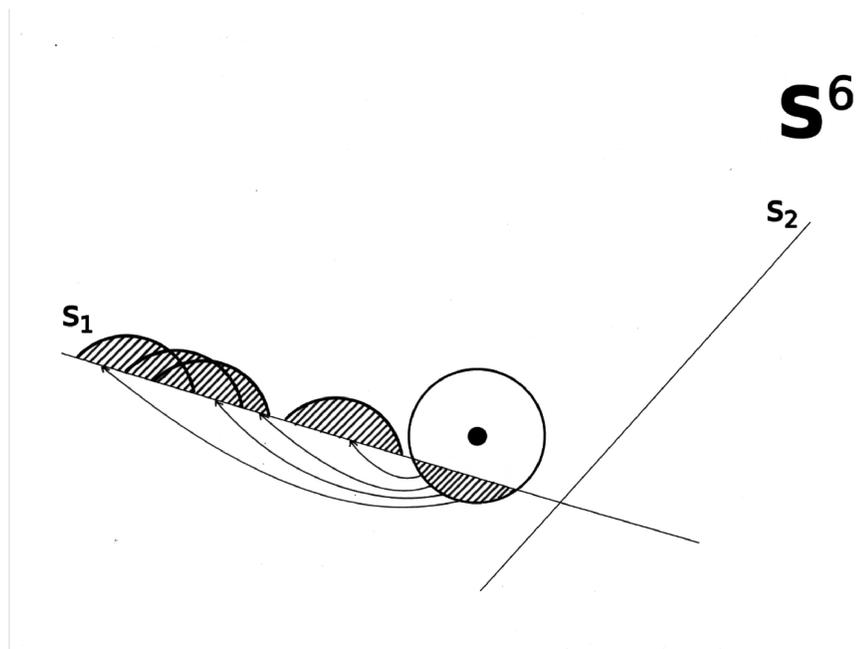}
	\caption{	Unlike in the 4-fold example, there are multiple 
		(up to 4) 
		places to go. A further complication is that the transformations
		at the boundaries do \textbf{not} preserve distances. Distances
		must be measured in the frame of the fundamental unit.}
	\label{S6_6}
\end{figure}

\begin{figure}

\includegraphics[height=0.6\textwidth]{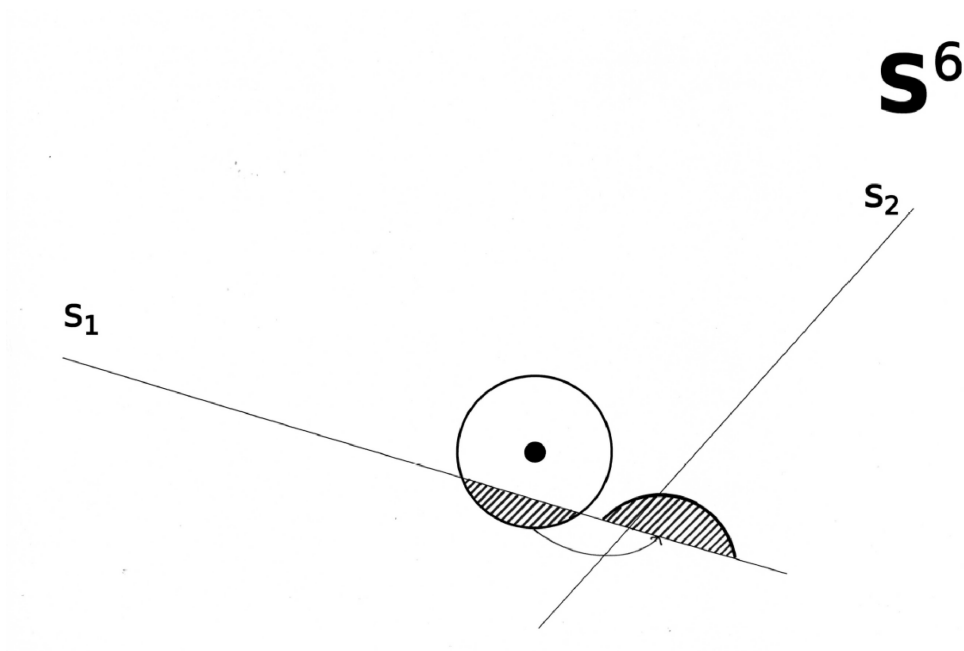}
\caption{Sometimes, in the process, another boundary is crossed. Then further iterations may be required.}
\label{S6_7}
\end{figure}

The distance between two lattices is determined by first computing $d_0$, the minimum of all the Euclidean distances between reflections of those two points within the fundamental unit.  If either lattice is closer to any boundary than $d_0$, then we need to iterate through all possible paths that may cross a boundary.

If the second point is near a boundary, then the four transforms for that boundary are applied. We need to compute the tunneled distances. If the generated points are not near another boundary, then we are done. If they are near a new boundary, then we need to iteratively apply the boundary transforms.

\begin{enumerate}
	\item Reduce both unit cells.
	\item Choose one point (``first point") to be the fixed point, and the other (``second point") to be the one to which transformations are applied.
	\item Compute $d_0$, the minimum of all the Euclidean distances between the first point and all reflections of the second point (the reflected points are all within the fundamental unit).
	\item Compute $d_1$, the minimum of all the Euclidean distances between the first point and all boundaries.
	\item Compute $d_2$, the minimum of all the Euclidean distances between the second point and all boundaries.
	\item If $d_1+d_2 < d_0$, then there may be a path between
	some reflection of the first point and  some reflection of
	the second point that is shorter than $d_0$ that crosses a boundary.  It is sufficient to explore possibly shorter
	paths by leaving the
	first point fixed and only considering reflections of the
	second point.  For each boundary to be considered there are
	at most four possible paths from whichever of the two points
	is close enough to that boundary to satisfy $d_1+d_2 < d_0$.
	For such a case we need to iterate through all four possible paths that may cross the boundary.
	\item If the second point is near two or three boundaries, then each boundary needs to be treated by itself. Then all pairs (and all triples) in all permuted orders must be transformed through the boundaries by all four paths of each boundary.  For a given \SVI{} boundary there are four possible boundary transforms to consider which may take a reflection
	of the second point outside of the fundamental unit (see Figure \ref{S6_6}).
	\item If the points produced in the above steps are near new boundaries, then the above steps need to be iterated using the new points. 
	\item N.B., the distance must be measured in the metric
	of the fundamental unit; see section \ref{measure}.
\end{enumerate}

In the above discussion of gluing in topology, mention was made of the
increase in dimensions as edges are glued and more if there are twists.
Here, the boundary transformation operations interchange two scalars, and
that is a twist operation. Counting up the gluings and twists leads to
the conclusion that the required dimension of the space in which to embed the glued manifold is quite high; immediately,
we understand that there are potentially going to be many paths to examine.

\subsection{Measuring with the metric of the fundamental unit}
\label{measure}
The problem that arises in computing distances between lattices
is that it is necessary to transform points to be outside the 
fundamental unit. The boundary transformations are not
isometric; distance measure is not preserved on crossing out
of the fundamental unit. Here we describe how to create a measure
for those points that is isometric.

\begin{figure}
	\includegraphics[width=.9\textwidth]{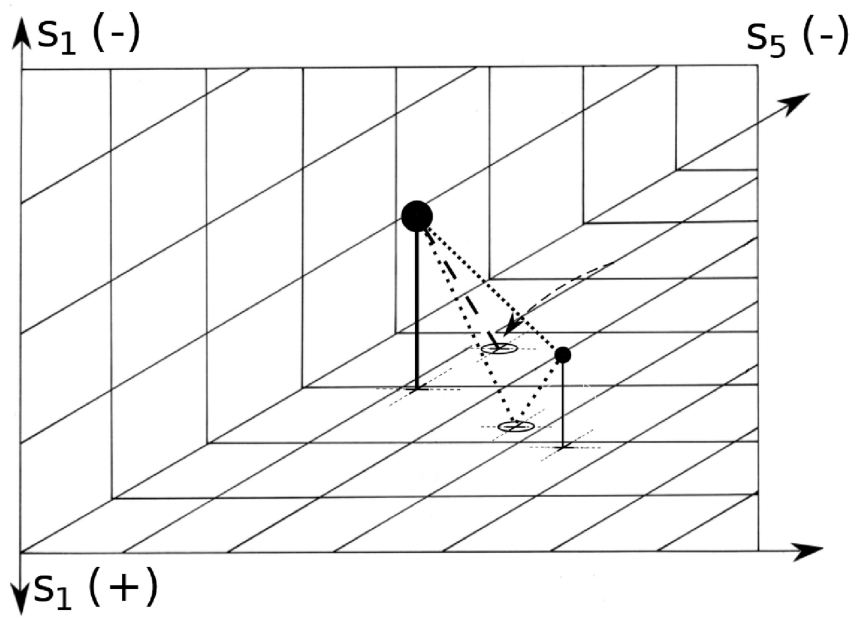}
\label{TMB}
	\caption{Tunneled Mirrored Boundary}
\end{figure}

If the second point of a measurement is near a boundary, then project the point
onto the boundary (the dashed arrow in Figure \ref{TMB}) and
extend the line segment from the second point to the boundary
to twice the length creating a transformed point creating
an artificial mirror point. The ``Tunneled Mirrored Boundaries" distance is found by locating the crossing point on the boundary where
the line from the first point to the artificial mirror point
crosses the boundary and then finding the four boundary transforms
of the crossing point.
Applying a boundary transform to
the crossing point provides an alternate boundary point
to be used for a broken path (first point to crossing point
point -- equivalent to boundary-mapped crossing point, followed
by mapped crossing point to second point).  This broken
path is sometimes shorter than the direct $d_0$ path and
is entirely in the fundamental unit.  If two boundaries
are close, then similar actions on both lattice points are 
needed.

\subsection{Algorithm}
	
	For finding the shortest distance between two reduced lattices, the first step
	is to compute the simple Euclidean distance between them.
	Each lattice is considered as a point in a space of six or
	more dimensions with a metric derived from the cell reduction
	rules for that space.
	
	If the original second point is near one (or more) boundaries, then the 
	the boundary searches (above) are performed. 
	
	If new near-boundary-contacts are detected, then the same kind of searches
	is performed.
	
	Distances must be measured in the metric of the fundamental unit 
	\textit{not in the metric of any region outside the fundamental unit}.

	\subsection{Nearest neighbors, the ``Post Office Problem"}
	
	Up to this point, the discussion has targeted finding the shortest 
	distance between two lattices. Sometimes, the problem is to find the collection of points
	that are near a particular probe. One important use is for
	creating and for searching within
	databases of crystals. Known as the nearest neighbor problem, it is also
	known as the post-office problem \cite{Andrews2001}, and \cite{Andrews2016}. (see Appendix \ref{nearest}.)

\section{Determining Bravais lattice types}

\subsection{Notation}

\begin{figure}
		\includegraphics[width=0.3\textwidth]{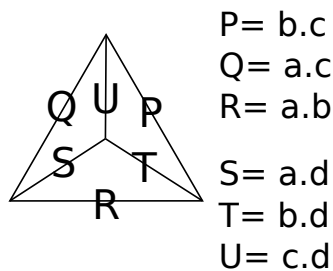} 
	\label{PQRSTU}
	\caption{Bravais organized the scalars into a ``tetrahedron"\cite{Delaunay1932}.}
\end{figure}
   The International
		Tables for Crystallography define this assignment of the scalars (PQRSTU) in
		the tetrahedron \cite{donnay1953international}. Other authors used alternative assignments.

		As \SVI{}: \textbf{s} = $\begin{pmatrix}{P,Q,R, S,T,U}\end{pmatrix}$\\ = $\begin{pmatrix}{s_1, s_2, s_3, s_4, s_5, s_6}\end{pmatrix}$.\\

		As \CIII{}:
		$\begin{bmatrix}
			\begin{pmatrix}
				P   \\
				S
			\end{pmatrix}
			\begin{pmatrix}
				Q   \\
				T
			\end{pmatrix}
			\begin{pmatrix}
				R   \\
				U
			\end{pmatrix}
		\end{bmatrix}$

	\subsection{Bravais lattice types in \SVI{}}
	
\begin{figure}
		\includegraphics[width=0.75\textwidth]{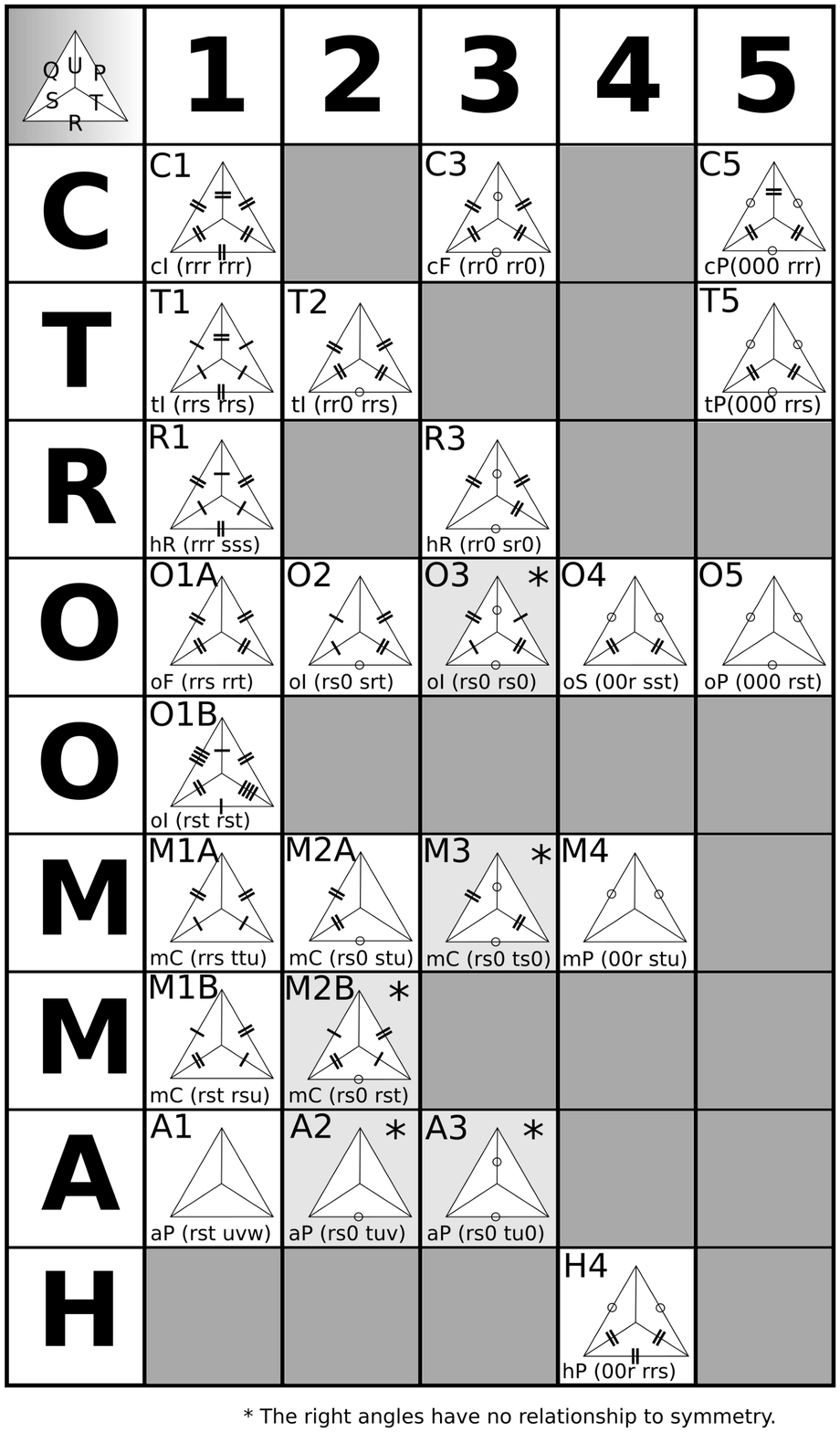}
		\caption{The table of \citeasnoun{Delaunay1932} describing the 24
			Bravais types in \SVI{}. It has been redone removing the images of the 
			Dirichlet cells, the non-reduced cells, and adding the ``lattice character", which
			describes the linear manifold of each type. The crystal family types have been
			renamed to modern usage: Q changed to T for tetrahedral, K changed
			to C for cubic, and T changed to A for anorthic. Where Delone
			in some places included two types in one table cell, they have been
			split into two (for example: ``M1" is split into ``M1A" and ``M1B".  Note that five types (O3, M3, M2B, A2, and A3) are not normal 
			crystallographic types. They are boundary types, and they have
			fewer free parameters than the generic type requires. For instance,
			O3 (character: rs0 rs0) has only two free parameters (r and s), whereas
			an ordinary orthorhombic type requires three variables.}
		\label{s6table}		
	\end{figure}

	\citeasnoun{Delaunay1932} and \citeasnoun{Delone1975} produced
	a table classifying the reduced forms of the Bravais types as 24 cases.
	Figure \ref{s6table} shows the same classification (the 
	letter codes for crystal systems have been changed to modern notation. For some types, revised names have been given
	 to Delone's types where a cell included two types (for example,
	Delone's ``M1" is split into ``M1A" and ``M1B"). In Figure \ref{s6table}, the 
	Bravais type is listed, and the restrictions on \SVI{} scalars is shown
	graphically as the Bravais tetrahedron and listed as the lattice character.

\begin{figure}
		\begin{center}
			\includegraphics[width=0.4\textwidth]{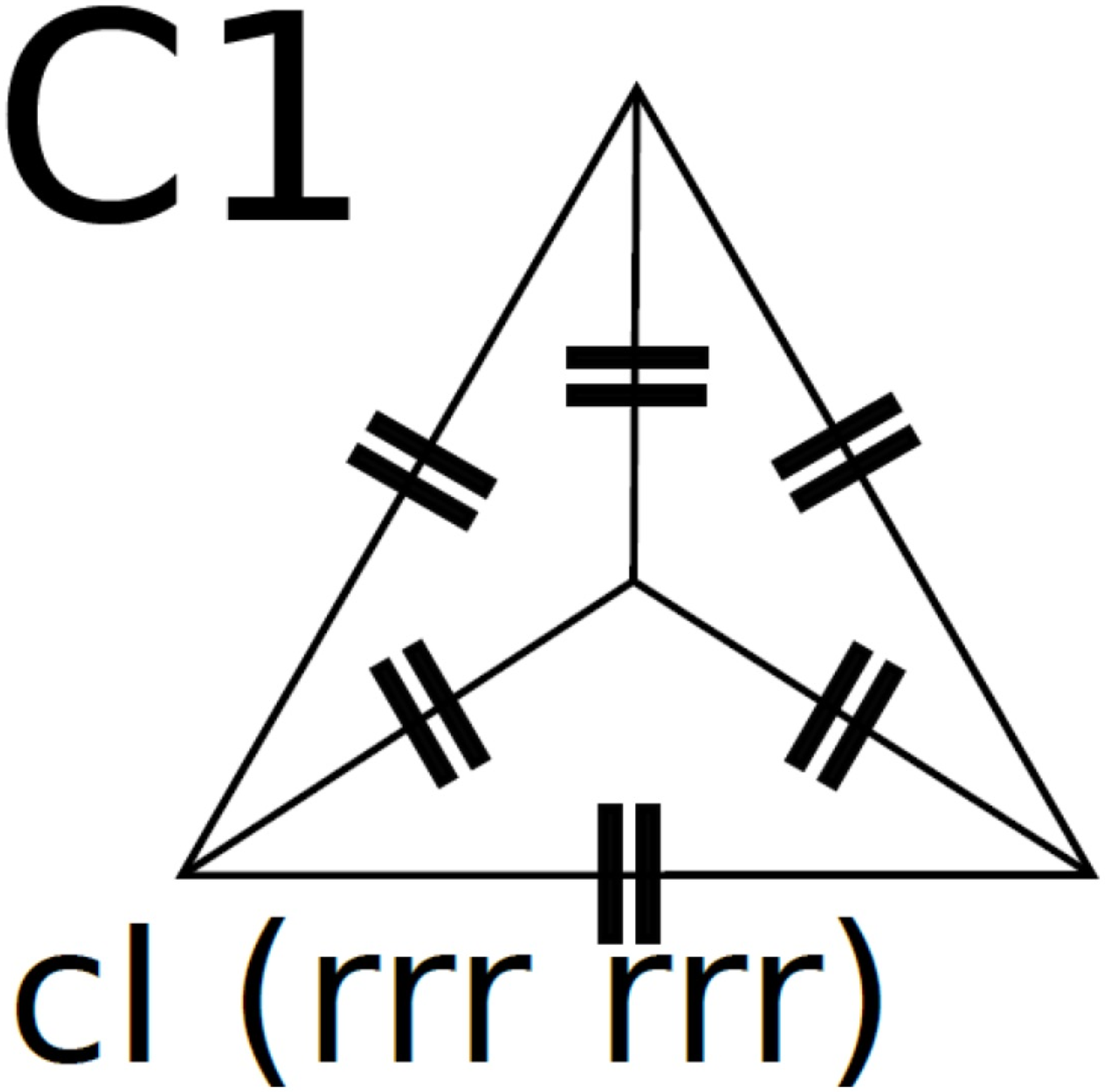}
			\includegraphics[width=0.4\textwidth]{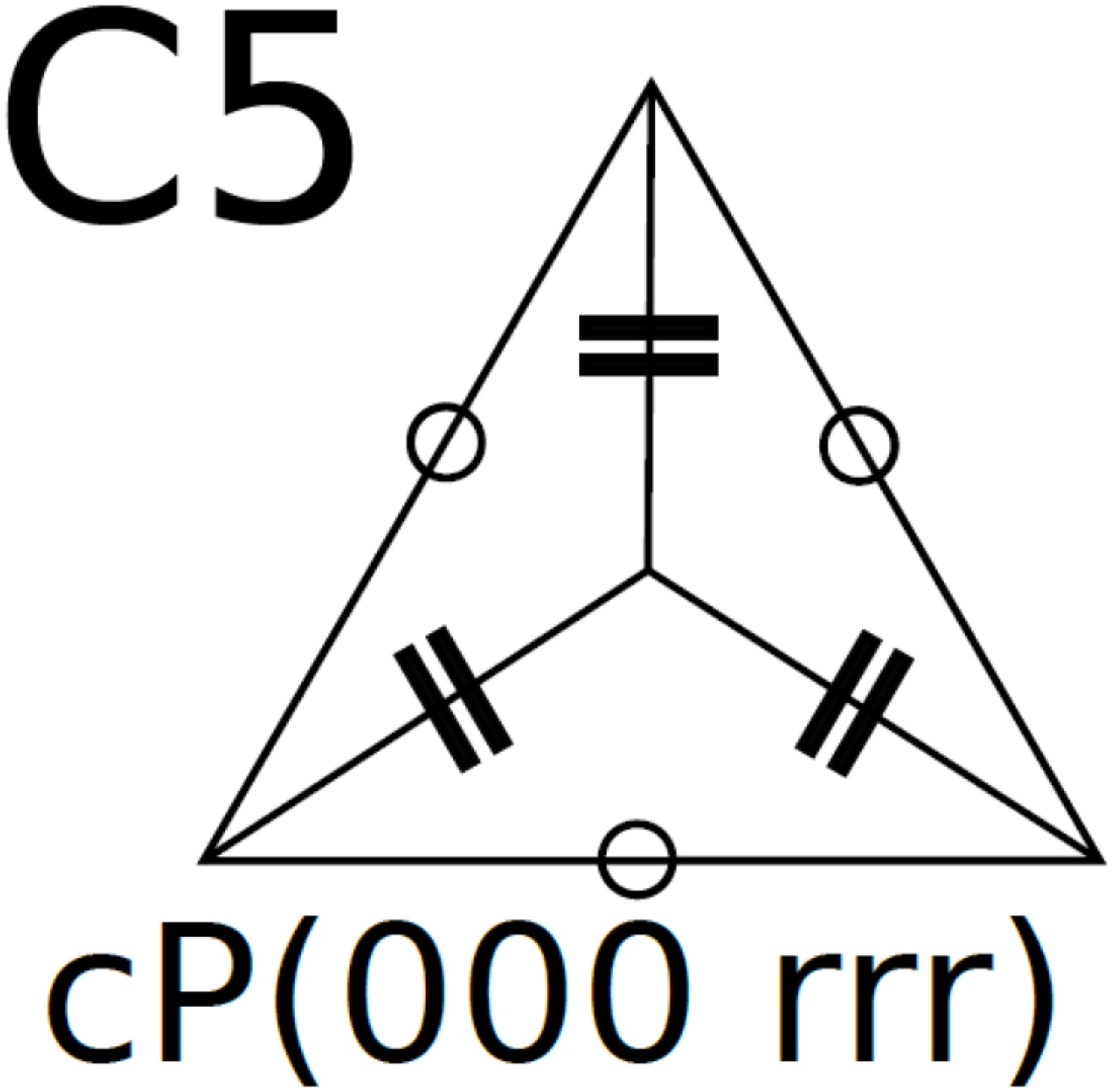}
			\label{Btype}
			\caption{Explaining the information in Figure \ref{s6table}; it includes the
				name of the Delone type (\textit{e.g.}: C1),the Bravais type
				(\textit{e.g.}: cI for body-centered cubic), and the lattice character that describes
				the linear manifold of each Delone type (\textit{e.g.}: 
				\mbox{000 rrr}).
				The markings on the tetrahedra describe the restriction on the \SVI{}
				scalars; they encode the same information that is in the lattice
				character. A zero (circle) means that the value of that scalar is
				zero, several with the same lines means they are constrained to be
				equal; in Figure \ref{Btype}, they are a pair of double
				line hatches.}
			
		\end{center}
		
	\end{figure}

	Figure \ref{Btype} describes the contents of each table entry
	in Figure \ref{s6table}.

\subsection{How to determine Bravais lattice types}
			
			First, the projectors for each type are necessary.  There are two ways to determine the Bravais lattice type.
			\begin{enumerate}
				\item The program BGAOL uses methods essentially the same as those used for distance calculations. 
				For a given input unit cell, the reflections and boundary transformations are applied. The 
				original cell and each of the transformed cells, is projected onto each of
				the manifolds of the lattice types (24 in the cases of \SVI{}). The distance from
				the probe cell to the projected cell is the agreement factor. For each lattice type,
				the cell with the shortest distance is reported.
				\item The program SELLA \cite{andrews2022unnecessarily} starts from vectors representing each of the lattice
				types and transforms them using the reflections and first layer boundary transformations.
				(In the case of multiple boundaries, all combinations must be used as in distance calculations).
				Then the distance from the reduced probe to each of the transformed manifolds of each type
				is computed. The shortest distance to one of the transformed manifolds is reported. If
				needed, the inverse reduction of the input (unreduced) unit cell can be applied to the 
				best projected cell to get the best approximation of the input.
				
			\end{enumerate}
	
	\subsection{What do you need to compute distances?}
	
\begin{itemize}
	\item 		Reflections
	 		\begin{itemize}
		 			\item 	\GVI{} \cite{Andrews2014}
		 			\item \SVI{} \cite{Andrews2019b}
		 		\end{itemize}
	\item	Boundary transforms
	 		\begin{itemize}
		 			\item 	\GVI{} \cite{Andrews2014}
		 			\item 	\SVI{} \cite{Andrews2019a} (only two, following Delone, but reflections generate a total of 4)
		 		\end{itemize}
	\item 		Bravais lattice projectors
	 		\begin{itemize}
		 			\item 	\GVI{} \cite{paciorek1992projection}
		 			\item 	\SVI{} \cite{andrews2022generating}
		 		\end{itemize}
	\item 		Lattice characters
	 		\begin{itemize}
		 			\item \GVI{} \cite{Andrews2014}
		 			\item \SVI{} see Section \ref{characters}
		 	\end{itemize}
\end{itemize}
	
\subsection{\SVI{} Bravais lattice characters}
		\label{characters}

Each Bravais type defined by \citeasnoun{Delaunay1932} corresponds
to a linear manifold in \SVI{}. The definitions of each type
were described by markings on the elements of the tetrahedra
for each type in Figure \ref{s6table}. For implementation
in software, it is
more convenient to use the ``lattice characters" that describe
the restrictions.

\begin{center}
	$\begin{tabular}  {lll lll lll lll}

		\begin{tabular}{l l}
			type & Bravais type \\
&\& character \\
			C1&cI (rrr rrr) \\ 
			C3&cF (rr0 rr0) \\ 
			C5&cP (000 rrr) \\ 
			T1&tI (rrs rrs) \\ 
			T2&tI (rr0 rrs) \\ 
			T5&tP (000 rrs) \\ 
			R1&hR (rrr sss) \\ 
			R3&hR (rr0 sr0) \\ 
			H4&hP (00r rrs) 
		\end{tabular}&&
	
		\begin{tabular}{l l}
			type & Bravais type \\
			&\& character \\
			O1A& oF (rrs rrt)               	  \\
			O1B& oI (rst rst)              	  \\
			O2&oI (rs0 srt)               	  \\
			\textit{O3} &\textit{oI} \textit{(rs0 rs0)} \textit{*}	\\
			O4&oS (00r sst)               	  \\
			O5&oP (000 rst)               	  \\
			O4&oS (00r sst)               	  \\
			O5&oP (000 rst)               	  \\
			&\\
		\end{tabular}&&
	
		$\begin{tabular}  {ll}       
			type & Bravais type \\
&\& character \\
			M1B &mC (rst rsu) \\
			M2A &mC (rs0 stu) \\
			\textit{M2B}& \textit{mC}  \textit{(rs0 rst)} {  *} \\
			\textit{M3}& \textit{mC}  \textit{(rs0 ts0)} {  *} \\
			M4 &mP  (00r stu) \\
			A1 &aP  (rst uvw) \\
			\textit{A2}& \textit{aP}   \textit{(rs0 tuv)} {  *} \\
			A3 &aP (rs0 tu0) {  *} \\
			M1A& mC (rrs ttu) 
		\end{tabular} $

	\end{tabular} $\\
{ *} cases of non-crystallographic \SVI{} scalars ($90^{\circ}$ angles)
\end{center}

\subsection{Lattice Matching}

A common problem in crystallography is to provide a list of the
unit cells of several (or many) crystals so that they can be visually
compared, making it easier to identify meaningful clusters
of crystals of related morphology. Data collections of
unit cells have been created based on similarity of morphology
(for example, see \citeasnoun{Donnay1963}.  In recent years,
the clustering of unit cells from the myriad of images in serial
crystallography has been increasingly important.

For each transformed vector, one must store the transformation that will bring it back to the original presentation.

Start with a collection of experimental unit cells. From
among them, we select or create the “reference” cell; that is the
one to which all the rest will be matched as closely as possible.

We transform the reference cell by many operations in the
course of exploring alternative lattice representations. For each
newly generated lattice representation,we accumulate the transformations
needed to convert back to the original reference cell.
All of these operations are be performed in \SVI{}.

Only accumulate transformations
that have not already been found (to avoid duplication).

Next all the accumulated, transformed representations of the
reference cell must be scaled  to a common length and the saved transformations
inverted which allows one to return a vector to
its original presentation.  For improved searching
in the final step, it is convenient to store the  the vectors in a nearest neighbor search
application such as NearTree \cite{Andrews2016}. This ends the 
preprocessing of the reference lattice.

To match a probe lattice to the reference cell, first the 
probe lattice should be
Selling reduced. Then the nearest point of the
collection of transformed lattice points to the reduced probe
is found. The stored transform of the nearest reference is
applied to the reduced probe to generate the best approximation.

As an example, consider the case of krait toxin phospholipase A2
\cite{LeTrong2007}. Table \ref{table:LT1} shows 6 reported forms
from the Protein Data Bank. The lattice type and unit cell parameters
are as initially reported. 

Table \ref{table:LT2} shows the same data in the same order after 
lattice matching to the first entry (1DPY). The immediate conclusion
is that the crystals are likely the same crystal form.

\begin{table}
	\begin{center}
		\caption{Unit cells of phospholipase A2 from the Protein Data Bank.}		
		\vspace{3mm}
		\begin{tabular}{lccccccccccc}
			\toprule
			PDB &   & a &b &c&$\alpha$&$\beta$&$\gamma$ \\
			\midrule
			1DPY &R& 57.98& 57.98 &57.98 &92.02  &92.02 &92.02 \\ 
			1FE5 &R& 57.98& 57.98 &57.98 &92.02  &92.02 &92.02 \\ 
			1G0Z &H& 80.36& 80.36 &99.44 &90     &90    &120   \\ 
			1G2X &C& 80.95& 80.57 &57.10 &90     &90.35 &90    \\ 
			1U4J &H& 80.36& 80.36 &99.44 &90     &90    &120   \\ 
			2OSN &R& 57.10 & 57.10  &57.10  &89.75  &89.75 &89.75  \\ 
			\bottomrule
		\end{tabular}
		\label{table:LT1}
	\end{center}
\end{table}

\begin{table}
	\begin{center}
		\caption{The data of Table \ref{table:LT1} matching a rhombohedral reference. The 
			reference cell is marked in boldface.}
		\vspace{3mm}
		\begin{tabular}{lccccccccccc}
			\toprule
			PDB &   a & b & c & $\alpha$ & $\beta$ & $\gamma$ \\ \midrule
			\textbf{1DPY} & \ \textbf{57.98}  & \textbf{57.98}  & \textbf{57.98}  &
			\textbf{92.02}  & \textbf{92.02}  & \textbf{92.02}  \\ 
			1FE5 &57.98  &57.98  &57.98  &92.02 & 92.02  & 92.02  \\ 
			1G0Z &57.02  &57.02  &57.02  &90.39  &90.39 &89.61  \\
			1G2X &57.10  &57.11  &57.11  &89.25  &90.25  &90.25  \\
			1U4J &57.02  &57.02  &57.02  &90.39  &90.39  &89.61  \\ 
			2OSN &57.10  &57.10  &57.10  &90.25  &90.25  &89.75  \\ 
			\bottomrule
		\end{tabular}		
		\label{table:LT2}
	\end{center}
\end{table}

%
\section{Various available programs}

	\itemize{
		\item SAUC: Search for related unit cells
		\item BGAOL: Determine likely cells of higher symmetry
		\item Follower: Compute distances between cells along a path
		\item LatticeMatching: Try to match a reference cell
		\item SELLA: Determine likely cells of higher symmetry
		\item various cmd programs\textendash
	}
\subsection{SAUC (Search of Alternative Unit Cells)}

	\textbf{SAUC} is software plus an online database of the Protein Data Bank unit cells and
the Crystallographic Open Database unit cells. \\
\url{"https://iterate.sourceforge.net/sauc/"}\\
SAUC provides searching for several alternate distance measures,
including \SVI{} and \GVI{}. Searching by nearest neighbor
allows specifying various criteria, such as percent range, 
nearest neighbors, or within a sphere of specified radius.\\

A useful example is that of krait toxin phospholipase A2
\cite{LeTrong2007}.

Starting from PDB entry 1u4j with cell parameters: (80.36, 80.36, 99.44, 90, 90, 120)	in space group R3 with a distance cutoff
of 3.5\AA  ~using the \GVI{} metric, four structures of phospholipase A2 are found.
\vspace{.2cm}

$\begin{tabular}{l l l}
	PDB code & G6 distance & protein name \\
	1u4j &  0.0 & {Phospholipase~A2~isoform 2} \\
	1g0z & 0.0 & Phospholipase~A2 \\
	1g2x & 0.9 &  Phospholipase~A2 \\
	2osn  & 0.9  & Phospholipase~A2~isoform 3 
	\label{phospholipase}
	\caption{}
\end{tabular}$

\vspace{.2cm}

\subsection{BGAOL - Bravais General Analysis of Lattices}

	\textbf{BGAOL} is an online program for seeking possible Bravais lattice types for a given  unit cell.
	
	\url{https://iterate.sourceforge.net/bgaol/} \\
	
  Consider 1,8-terpin \cite{Mighell2001} with unit cell 
10.912(2), 22.79(4), 10.705(2), 90, 120.64, 90.

The BGAOL results (only the best 4, which are also 1,8-terpins) are:

\begingroup
\setlength{\tabcolsep}{1.6pt}
\renewcommand{\arraystretch}{0.925}
\begin{tabular}{llcrlcccccc}
 lattice &  distance &&unit& cell\\
 oF  &  \bf{0.025}&18.419 &22.790 &10.912& 90& 90 & 90 \\
mC  &  \bf{0.025}& 10.912&18.419  &12.634 &90&115.585 &90\\
mI&\bf{0.000}&10.704 &22.790&10.705&90&118.713& 90\\
 mI  &  \bf{0.018} &14.652 &10.912& 14.651&90&102.109 &90\\
	\caption{}
	\label{bgaol}
\end{tabular}
\endgroup

The unit cells reported in Table \ref{bgaol} are then
matched to the lattice type in column 1. If the 
discovered lattice type is monoclinic (in this case mI),
then the unit cell is projected into the manifold of the 
discovered type.

A web site on which searches may be done  is located at \url{http://iterate.sourceforge.net/bgaol/}.

\section{Summary}

Unit cells are used to represent crystallographic lattices.
Calculations measuring the differences between unit
cells are used to provide metrics for measuring meaningful
distances between three-dimensional crystallographic
lattices.  This is a surprisingly complex and
computationally demanding problem.  We present a
review of the current best practice using 
Delaunay-reduced unit cells
in the six-dimensional real space of Selling
scalar cells \SVI{} and the equivalent three-dimensional complex space \CIII{}.
The process is a simplified version of the process
needed when working with the more complex
six-dimensional real space of Niggli-reduced unit
cells \GVI{}.  Obtaining a distance begins with identification of the fundamental region in the space, continues with
conversion to primitive cells and reduction, analysis of
distances to the boundaries of the fundamental unit,
and is completed by a comparison of direct paths to
boundary-interrupted paths, looking for a path of
minimal length. 

\appendix
\section{Availability of code}
    \label{code}

Code supporting this paper is available on github
at \url{http://github.com/duck20/LatticeRepLib.git}
and \url{http://github.com/yayahjb/ncdist.git}.

\section{Projectors}
	\label{projectors}
	
Orthogonal projectors onto linear subspaces can be calculated from 
any algebraic descriptor of that subspace that can provide a matrix $X$ with vectors that at least span that subspace.  Redundant vectors can help to improve the accuracy of the calculation.  If $A$ is the columnspace of $X$, then
$P=A(A^TA)^{-1}A^T$ is a projector onto that linear subspace,
and $PP=P$ and $P^T=P$.

The application of a projector yields the least-squares best fit.

\begin{figure}
				\includegraphics[width=0.8\textwidth]{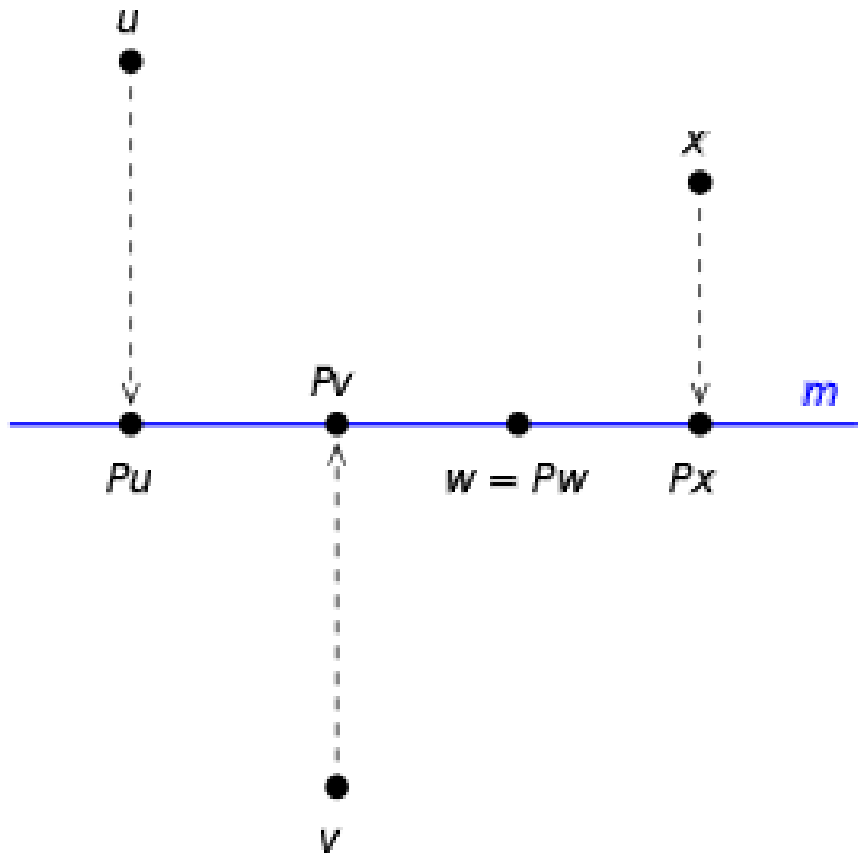}

	\caption{Example of projectors: they are the operators that do orthogonal projection.}
\end{figure}
\section{Nearest neighbor search}
\label{nearest}

Wikipedia has a useful description of nearest neighbor searching \cite{enwiki:1103897119}.

Code for nearest neighbor searching using NearTree is available on GitHub
at \url{http://github.com/yayahjb/neartree.git}. NearTree uses the algorithm of
\citeasnoun{kalantari1983data} with enhancements for other kinds of searches. 

\section{Polar coordinates}
\label{polar}

Expressing unit cell parameters as edge lengths and angles does 
not lead to a easily interpreted metric: lengths and angles are
not commensurate. An unexplored alternative is to express them in
polar coordinates, which is equivalent to expressing them in complex
coordinates. 

Delone pointed
out that the ``opposite" scalars form special pairs. The Bravais
tetrahedron is the origin of his use of the term ``opposite". Another way
to view this relationship is to organize the six scalars into
three pairs, treating the first member of the each pair as
the real component of a complex number and the second member of
each pair as the imaginary component.
Write the definition of an \SVI{} vector $(s_1, s_2, s_3, s_4, s_5, s_6)$ in \CIII{}:\\

$\begin{bmatrix}$
	\begin{tabular}{ c c l c }
		$\begin{pmatrix}
			s_1 \\
			s_4 
		\end{pmatrix}$ 
		
		$\begin{pmatrix}
			s_2 \\
			s_5
		\end{pmatrix}$
		
		$\begin{pmatrix}
			s_3\\
			s_6
		\end{pmatrix}$
	\end{tabular}
	$\end{bmatrix}$ \\ \\

Rewriting the definitions of the \SVI{} scalars, we
see that for each \CIII{} scalar, the angle in the real
position, is related to the edge vector used in the 
imaginary position. For instance for $c_1$, alpha is 
associated with $a$. For reflections, these associations
are not changed.\\

$\begin{bmatrix}$
	\begin{tabular}{ c c l c }
	$\begin{pmatrix}
		\bf{b \cos{\alpha}~ c} \\
		\bf{a \cdot d} 
	\end{pmatrix}$ 
	
	$\begin{pmatrix}
		\bf{a \cos{\beta} ~c} \\
		\bf{b \cdot d}
	\end{pmatrix}$
	
	$\begin{pmatrix}
		\bf{a \cos{\gamma}~ b}\\
		\bf{c \cdot d}
	\end{pmatrix}$
\end{tabular}
	$\end{bmatrix}$ \\

The stability of the angle/edge length association implies
that a potentially  useful new coordinate system for describing unit cells 
might be $\begin{pmatrix} a,\alpha \end{pmatrix}$, 
 $\begin{pmatrix} b,\beta \end{pmatrix}$,  
 $\begin{pmatrix} c,\gamma \end{pmatrix}$.

\section{The linear space \GVI ~(Niggli space)}
\label{G6}

			Label  the cell axes ${\textbf{a}, \textbf{b}, \textbf{c}}$. The scalars are	\vdotv{\textbf{a}}{\textbf{a}},~ \vdotv{\textbf{b}}{\textbf{b}},~ \vdotv{\textbf{c}}{\textbf{c}},
			~ 2\vdotv{\textbf{b}}{\textbf{c}},~ 2\vdotv{\textbf{a}}{\textbf{c}},~ 2\vdotv{\textbf{b}}{\textbf{c}} where,
			{\it e.g.},
			\vdotv{\textbf{b}}{\textbf{c}} represents the dot product of the ${\textbf{b}}$  and ${\textbf{c}}$  axes. In order to organize these six quantities as a vector space in which one
			can compute simple Euclidean distances, we describe this set of scalars 
			as a vector, ${\textbf{g}}$, with components, $ {g}_1, {g}_2, {g}_3, {g}_4, {g}_5, {g}_6 $. \cite{Andrews2014}
		It should be noted that the first three scalars of \GVI{} are the
		 diagonal elements of the metric tensor. The second three are the 
		 off-diagonal elements, doubled. One reason for doubling is that those
		 elements occur twice within the metric tensor.

\section{The linear space \SVI{}~ (Delone space)}
\label{S6}

		\SVI{}: \citeasnoun{Andrews2019b} 
			presented a simple and the fastest currently
			known representation of lattices as the six Selling 
			scalars obtained from the dot products of the unit cell axes in addition to the negative of their sum (a body diagonal). Labeling 
			the cell axes ${\textbf{a}, \textbf{b}, \textbf{c}}$ , and ${\textbf{d}}$ (${\textbf{d}} =$ $-{\textbf{a}}-\!{\textbf{b}}-\!{\textbf{c}}$), 
			The scalars are	\vdotv{\textbf{b}}{\textbf{c}},~ \vdotv{\textbf{a}}{\textbf{c}},~ \vdotv{\textbf{a}}{\textbf{b}},
			~ \vdotv{\textbf{a}}{\textbf{d}},~ \vdotv{\textbf{b}}{\textbf{d}},~ \vdotv{\textbf{c}}{\textbf{d}} where,
			{\it e.g.},
			\vdotv{\textbf{b}}{\textbf{c}} represents the dot product of the ${\textbf{b}}$  and ${\textbf{c}}$  axes. For the purpose of organizing these six quantities as a vector space in which one can compute simple Euclidean distances, we describe this set of scalars 
			as a vector, ${\textbf{s}}$, with components, $ {s}_1, {s}_2, {s}_3, {s}_4, {s}_5, {s}_6 $.\\
		 		{ A cell is Selling-reduced if all six components are negative or zero \cite{Delaunay1932}.  
			\item Reversing the  observation that a Buerger-reduced cell is a good stepping stone to a Selling-reduced cell \cite{Allmann1968},
			a Selling-reduced cell can be a very efficient stepping stone to a Buerger-reduced cell from which a Niggli-reduced cell can quickly be derived.
			\item {Selling showed that the all-obtuse (Selling parameters all-negative) case is unique.
			 Every \SVI{} vector with all scalars non-positive corresponds to a valid unit cell.}

\section{The linear space \CIII{}}
\label{C3}

	\CIII{}~ is a space created from the scalars of \SVI{}. \\
		\vspace{.1cm}
	
\begin{center}
		$\begin{bmatrix}
		\begin{pmatrix}
			s_1 \\
			s_4
		\end{pmatrix}
		\begin{pmatrix}
			s_2 \\
			s_5
		\end{pmatrix}
		\begin{pmatrix}
			s_3 \\
			s_6
		\end{pmatrix}
	\end{bmatrix}$\\
\end{center}	
    ~~\\
	This associates each axis with its associated angle \cite{Andrews2019b}. 
	~~\\
	
	\begin{center}
		$\begin{bmatrix}
			\begin{pmatrix}
				\alpha   \\
				\textbf{a}
			\end{pmatrix}
			\begin{pmatrix}
				\beta   \\
				\textbf{b}
			\end{pmatrix}
			\begin{pmatrix}
				\gamma   \\
				\textbf{c}
			\end{pmatrix}
		\end{bmatrix}$
	\end{center}

\section{Geodesics}
\label{geodesic}

A geodesic is a curve representing in some sense the shortest path between two points in a surface \cite{enwiki:1116672577}.

\ack{{\bf Acknowledgements}}

Careful copy-editing and corrections by Frances C. Bernstein are 
gratefully acknowledged. Elizabeth Kincaid is to be thanked
for artwork.

\ack{{\bf Funding information}}      

Funding for this research was provided in part by:  
US Department of Energy Offices of Biological and 
Environmental Research and of Basic Energy Sciences 
(grant No. DE-AC02-98CH10886; grant No. E-SC0012704); 
U.S. National Institutes of Health (grant No. P41RR012408; 
grant No. P41GM103473; grant No. P41GM111244; 
grant No. R01GM117126,
grant No. 1R21GM129570); Dectris, Ltd.

\bibliography{Reduced}

\bibliographystyle{iucr}




\end{document}